\documentclass[journal]{IEEEtran}
\usepackage{amsmath,amsfonts,amssymb}
\usepackage{algorithmic}
\usepackage{algorithm}
\usepackage{array}
\usepackage[caption=false,font=footnotesize,labelfont=rm,textfont=rm]{subfig}
\usepackage{bm}
\usepackage{textcomp}
\usepackage{stfloats}
\usepackage{url}
\usepackage{verbatim}
\usepackage{graphicx}
\usepackage{cite}
\usepackage{multirow}
\usepackage{booktabs}

\renewcommand{\arraystretch}{0.95}
\begin{document}

\title{Indirect and Direct Multiuser Hybrid Beamforming for Far-Field and Near-Field Communications: A Deep Learning Approach}

\author{Xinyang Li, Songjie Yang, Boyu Ning, Zongmiao He, Xiang Ling,~\IEEEmembership{Member,~IEEE,} and Chau Yuen,~\IEEEmembership{Fellow,~IEEE,}
\thanks{This work was supported in the National Science and Technology Major Project of China under Grant 2024ZD1300800.}
\thanks{Xinyang Li, Songjie Yang, Boyu Ning,  and Xiang Ling are with the National Key Laboratory of Wireless Communications, University of Electronic Science and Technology of China, Chengdu 611731, China (e-mail: lixy829@std.uestc.edu.cn; yangsongjie@std.uestc.edu.cn;   boydning@outlook.com; xiangling@uestc.edu.cn).
        
        Zongmiao He is with the School of Network \& Communication Engineering, Chengdu Technological University, Chengdu 611730, China, and also with the National Key Laboratory of Wireless Communications, University of Electronic Science and Technology of China, Chengdu 611731, China (email: hzmiao@cdtu.edu.cn).
        
		Chau Yuen is with the School of Electrical and Electronics Engineering, Nanyang Technological University (e-mail: chau.yuen@ntu.edu.sg).
}
\thanks{This work has been submitted to the IEEE for possible publication. Copyright may be transferred without notice, after which this version may no longer be accessible.}}


\maketitle

\begin{abstract}
Hybrid beamforming for extremely large-scale multiple-input multiple-output (XL-MIMO) systems is challenging in the near field because the channel depends jointly on angle and distance, and the multiuser interference (MUI) is strong.
Existing deep learning methods typically follow either a decoupled design that optimizes analog beamforming without explicitly accounting for MUI, or an end-to-end (E2E) joint analog--digital optimization that can be unstable under nonconvex constant-modulus (CM), pronounced analog--digital coupling, and gradient pattern of sum-rate loss.
To address both issues, we develop a complex-valued E2E framework based on a variant minimum mean square error (variant-MMSE) criterion, where the digital precoder is eliminated in closed form via Karush--Kuhn--Tucker (KKT) conditions so that analog learning is trained with a stable objective.
The network employs a grouped complex-convolution sensing front-end for uplink (UL) measurements, a shared complex multi-layer perceptron (MLP) for per-user feature extraction, and a merged constant-modulus head to output the analog precoder.
In the indirect mode, the network designs hybrid beamformers from estimated channel state information (CSI).
In the direct mode where explicit CSI is unavailable, the network learns the sensing operator and the analog mapping from short pilots, after which additional pilots estimate the equivalent channel and enable a KKT closed-form digital precoder.
Simulations show that the indirect mode approaches the performance of iterative variant-MMSE optimization with a complexity reduction proportional to the antenna number.
In the direct mode, the proposed method improves spectral efficiency over sparse-recovery pipelines and recent deep learning baselines under the same pilot budget.
\end{abstract}


\section{Introduction}

\IEEEPARstart{T}{he} rapid growth of global mobile data traffic necessitates higher spectral efficiency in wireless networks \cite{ITU-R_M.2370-0, Ericsson_MobileDataTrafficOutlook}. 
While fifth-generation (5G) technologies, such as millimeter-wave communications and massive multiple-input multiple-output (mMIMO), have improved spectral efficiency \cite{itu2017minimum}, sixth-generation (6G) networks require further gains in peak data rates. 
This motivates the use of higher-frequency bands with wider bandwidths and denser antenna deployments.
This trend has promoted \emph{extremely large-scale antenna arrays (ELAAs)} and the associated \emph{extremely large-scale MIMO (XL-MIMO)} architectures. 
In these systems, thousands of elements enable narrow beams to meet the ultra-high throughput and low-latency requirements of 6G networks \cite{saad2019vision, wang2024tutorial}.

The large aperture of ELAAs extends the Rayleigh distance, placing users in the near field \cite{10934792, an2024near, chen2024unifying}. 
As a result, the planar-wave assumption becomes inaccurate and a spherical-wave model is needed, where distance explicitly enters the array response \cite{10500431, gong2024holographic, gong2024near}. 
Although this additional distance dimension introduces model mismatch for far-field algorithms, it enables \emph{beam focusing}---a precise spatial multiplexing technique based on joint angle--distance beamforming \cite{cui2022channel,11017428}. 
Accordingly, \emph{Location Division Multiple Access (LDMA)} enabled by beam focusing can separate users with similar angles but different ranges, unlike conventional far-field \emph{Space Division Multiple Access} \cite{wu2023multiple}. 
However, optimizing hybrid beamformers in this high-dimensional space to maximize the sum rate and suppress multiuser interference (MUI) is nonconvex and computationally intensive. 
Deep learning offers an efficient surrogate for the resulting high-dimensional nonconvex mapping and has been increasingly considered in artificial intelligence (AI)-native radio design. 

\subsection{Prior Works}\label{SecIA}
In conventional far-field mMIMO, large arrays generate high-resolution \emph{pencil beams}. 
However, fully digital beamforming requires numerous radio frequency (RF) chains, resulting in high hardware complexity and power consumption \cite{gao2016energy, sohrabi2016hybrid}. 
Hybrid architectures address this challenge by combining limited RF chains with analog phase shifters. 
A common two-stage framework identifies propagation directions via exhaustive scanning \cite{gao2016energy} or hierarchical codebook search \cite{xiao2016hierarchical, li2023hierarchical}. 
It constructs analog precoders to compensate for path loss and then designs digital precoders based on equivalent baseband channels \cite{park2017exploiting}. 
Although hierarchical schemes reduce pilot overhead, they face four main limitations: i) excessive feedback requirements \cite{qi2020hierarchical}; ii) limited angular resolution; iii) reliance on the planar-wave assumption; and iv) error propagation across layers. 
These limitations compromise beamforming precision in near-field scenarios, where focusing energy over specific ranges and angles is essential.

\IEEEpubidadjcol
Alternative approaches approximate fully digital solutions by minimizing the Euclidean distance between hybrid and unconstrained precoders \cite{el2014spatially, yu2016alternating} or by reformulating sum rate maximization via weighted minimum mean square error (WMMSE) criterion \cite{sun2016majorization, christensen2008weighted,2025arXiv250317718Y}. 
These methods typically entail high computational complexity that scales quadratically or cubically with the number of antennas. 
They also assume perfect channel state information (CSI), which is unrealistic in ELAA systems due to channel estimation overhead \cite{gao2016channel}. 
In near-field conditions, channel orthogonality breaks down, making joint beamforming optimization more difficult.

Near-field propagation enables spherical-wave beam focusing, which increases the search complexity in high-dimensional angle--range space. 
In single-user scenarios, codebook-based methods have demonstrated effective pilot reduction \cite{lu2023hierarchical, zhang2022fast, liu2022deep}. 
In multiuser settings, overlapping beams lead to shared spatial regions, reducing the orthogonality of analog precoders, and signal leakage between users, which causes significant residual interference after hybrid beamforming \cite{wu2023multiple}. 
Although dynamic metasurfaces \cite{zhang2022beam, yang2025beam} offer programmable focusing, their non-uniform and angle-dependent sidelobes \cite{wang2019dynamic} make resource scheduling and practical deployment difficult in dense networks.

Deep neural networks (DNNs) \cite{ma2021deep, liu2024near} reduce online computational complexity by learning offline mappings from CSI to precoders. 
Many existing approaches formulate beam training as a classification task using far-field discrete Fourier transform codebooks or their near-field extensions \cite{ma2021deep}.
However, the quantization inherent in these codebooks restricts the adaptability of beam focusing. 
While recent graph neural networks exploit user spatial correlations \cite{liu2024near}, they remain constrained by codebook resolution, which precludes continuous beam focusing. 
In addition to beamforming, deep learning has also been applied to pilot-based channel estimation in mMIMO systems \cite{chun2019deep,kang2020deep}, with these methods serving as upstream modules when explicit CSI acquisition is required. 
Building on learning-aided designs, solutions such as NMAP-Net \cite{kang2025nmap} have further extended near-field XL-MIMO optimization to include joint beamforming and position optimization.

End-to-end (E2E) architectures have gained attention for their versatility in both near- and far-field scenarios, operating without traditional codebooks. 
The authors of \cite{attiah2022deep} propose a two-stage DNN that decouples analog and digital components via a sum-rate reformulation. 
While this simplifies the design, it ignores the digital precoder's role in suppressing MUI during the analog optimization phase. 
In \cite{gao2022data}, residual networks are utilized for wideband beamforming.
Minimizing the negative sum rate often results in unstable gradients; combined with rectified linear unit (ReLU) saturation, this undermines the effectiveness of backpropagation. 
The framework in \cite{park2024end} integrates pilot adaptation and power allocation.
Nevertheless, its fully digital architecture leads to high power consumption and hardware costs, posing challenges for ELAA systems. 
The authors of \cite{nie2024near} adopted convolutional neural networks to generate interference-aware analog precoders but omitted digital precoding for residual interference suppression. 
Additionally, their definition of sum rate deviates from standard formulations. 
In the context of pinching antenna systems, \cite{xu2025joint} investigated joint transmit and pinching beamforming from both optimization and learning perspectives.
This motivates integrating Karush--Kuhn--Tucker (KKT) optimality conditions into learning-based designs.

\subsection{Main Contributions}
To address these limitations, we propose an E2E hybrid beamforming framework based on deep complex networks (DCNs) \cite{Trabelsi2017deep}.
The framework supports both indirect and direct designs.
In indirect mode, the network designs hybrid beamformers using estimated CSI.
In direct mode, without explicit CSI, it maps limited UL pilots to an optimal analog beamformer, then applies a closed-form digital precoder from equivalent baseband channel estimation.
We adopt a variant-MMSE-based training objective to stabilize the beamforming optimization.
Our main contributions are:
\begin{itemize}
\item We present a fully complex-valued E2E DNN for multiuser hybrid beamforming in both near- and far-field regimes. The architecture comprises a grouped complex-convolution sensing layer that emulates uplink (UL) measurements, a shared per-user complex multi-layer perceptron (MLP) with complex batch normalization and a custom complex hyperbolic tangent (Tanh) activation, and a merged output head with constant-modulus (CM) normalization to produce analog beamformers. This design preserves the amplitude--phase structure and coordinates user features to mitigate MUI.
\item We propose a data-driven framework for both indirect and direct hybrid beamforming, which extracts high-resolution angle--distance features from either perfect CSI (PCSI) or limited UL pilots, and generalizes well across near- and far-field channels. In indirect mode, the proposed method reduces computational complexity by an order proportional to the number of antennas while achieving near-optimal performance. In direct mode, it reduces pilot and feedback overhead, which improves scalability in XL-MIMO deployments.
\item We develop a variant-MMSE-based objective as a stable surrogate for sum-rate maximization, where the digital precoder is analytically eliminated via its closed-form KKT solution. This strategy effectively decouples the analog and digital designs, circumventing the gradient instability inherent in conventional E2E sum-rate optimization. Consequently, it ensures robust convergence and superior spectral efficiency in both indirect and direct beamforming modes.
\item Experiments show that the proposed framework outperforms representative baselines, achieving sum-rate gains of up to 3~bps/Hz with enhanced pilot efficiency and reduced computational complexity. Beyond performance metrics, we provide interpretability into the learning mechanism by visualizing the sensing patterns and latent feature manifolds, confirming that the network captures physically meaningful, range-dependent spatial structures.
\end{itemize}

{\it Organization:} Section~II presents the system model, including the UL and downlink (DL) signal models, near- and far-field channel characteristics, and the multiuser sum-rate formulation. Section~III details the proposed E2E architecture, the variant-MMSE-based objective, and the indirect beamforming algorithm using estimated channels. Section~IV introduces the direct hybrid beamforming approach for scenarios where explicit CSI is unavailable. Section~V presents simulation results to validate the proposed framework, followed by conclusions in Section~VI.

{\it Notations:} Bold uppercase and lowercase letters denote matrices and vectors; $(\cdot)^T$, $(\cdot)^H$ and $(\cdot)^*$ denote the transpose, Hermitian transpose, and conjugate operations; $\mathbf{I}_N$ represents the $N{\times}N$ identity matrix; $\|\cdot\|_F$ and $\|\cdot\|_2$ denote the Frobenius and Euclidean norms; $|\cdot|$ denotes the modulus operator; $\mathrm{tr}(\cdot)$ is the trace; $\mathfrak{R}(\cdot)$ and $\mathfrak{I}(\cdot)$ extract the real and imaginary parts; $\mathbb{E}[\cdot]$ is expectation; $\mathbb{C}$ and $\mathbb{R}_+$ are the sets of complex and positive real numbers; $\mathcal{CN}$ and $\mathcal{U}$ denote the complex Gaussian and uniform distributions. For a matrix $\mathbf{A}$, $[\mathbf{A}]_{i,j}$ denotes its $(i,j)$-th entry; for a 3-D tensor $\mathbf{A}$, $[\mathbf{A}]_{i,:,:}$ represents the slice along the first mode. 

\section{System Model}

\begin{figure}[!t]
\centering
\includegraphics[width=3.5in]{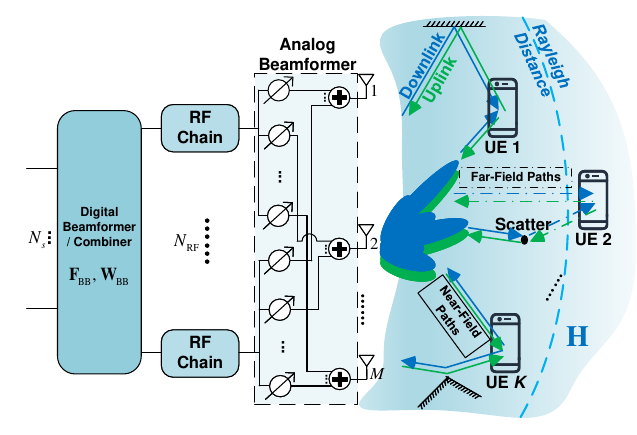}
\caption{Schematic of the XL-MISO hybrid beamforming system serving user equipments (UEs) in mixed near- and far-field regions.}
\label{fig:system}
\end{figure}
We consider a narrowband time-division duplex (TDD) multiuser multiple-input single-output (MU-MISO) system. 
As shown in Fig.~\ref{fig:system}, a base station (BS) with a uniform linear array (ULA) of $M$ antennas and $N_{\rm RF}$ RF chains uses hybrid precoding to serve $K$ single-antenna users.
We assume $K\le N_{\rm RF}\ll M$, which is typical for ELAA deployments with users distributed in both the near- and far-field regions.

During the UL pilot transmission phase under a time-division multiple-access (TDMA) protocol, each user $k \in \mathcal{K} = \{1, \dots, K\}$ transmits a pilot symbol $x_k$ with unit power, i.e., $\mathbb{E}[|x_k|^2] = 1$, in its assigned time slot. 
In the $n$-th sensing slot, the BS applies a fixed combining matrix ${\bf \Phi}^{(n)} \in \mathbb{C}^{K \times M}$ to process the pilot symbols received over $K$ consecutive user slots. 
This effective combiner can be realized by hybrid combining, where ${\bf \Phi}^{(n)} = {\bf W}_{\rm BB}^{(n)H}{\bf W}_{\rm RF}^{(n)H}$ and only ${\bf W}_{\rm RF}^{(n)}$ satisfies the CM constraint.
The combined UL signal ${\bf y}_{k, {\rm UL}}^{(n)} \in \mathbb{C}^{K \times 1}$ from user $k$ is given by
\begin{equation}
    {\bf y}_{k, {\rm UL}}^{(n)} = {\bf \Phi}^{(n)} \left( {\bf h}_{k} x_k + {\bf v}_{k, {\rm UL}}^{(n)} \right),
    \label{eq:ul_single_user}
\end{equation}
where ${\bf h}_{k} \in \mathbb{C}^{M \times 1}$ denotes the UL channel vector from user $k$, and ${\bf v}_{k, {\rm UL}}^{(n)} \sim \mathcal{CN}({\bf 0}, \sigma_n^2 {\bf I}_M)$ represents the independent and identically distributed (i.i.d.) additive white Gaussian noise (AWGN) vector. 
Following combining, the effective noise becomes colored with covariance $\sigma_n^2 {\bf\Phi}^{(n)}({\bf\Phi}^{(n)})^H$. 
Since the pilots are known and have unit power, we omit $x_k$ in the sequel without loss of generality.
By collecting the received signals from all $K$ users, we obtain the aggregate signal matrix ${\bf Y}_{\rm UL}^{(n)} \in \mathbb{C}^{K \times K}$:
\begin{equation}\begin{aligned}
    {\bf Y}_{\rm UL}^{(n)} \triangleq &\left[ {\bf y}_{1,{\rm UL}}^{(n)}, {\bf y}_{2,{\rm UL}}^{(n)}, \cdots, {\bf y}_{K,{\rm UL}}^{(n)} \right] \\= &{\bf \Phi}^{(n)} {\bf H}_{\rm UL} + {\bf \Phi}^{(n)} {\bf V}_{\rm UL}^{(n)},
    \label{eq:ul_slot}
\end{aligned}\end{equation}
where ${\bf H}_{\rm UL} \triangleq \left[{\bf h}_{1}, \dots, {\bf h}_{K}\right]$ aggregates the UL channel vectors, and ${\bf V}_{\rm UL}^{(n)} = \left[{\bf v}_{1, {\rm UL}}^{(n)}, \dots, {\bf v}_{K, {\rm UL}}^{(n)}\right]$ denotes the UL noise matrix. 
Stacking the measurements over $N$ consecutive sensing slots yields the global observation matrix ${\bf Y}_{\rm UL} \in \mathbb{C}^{N K \times K}$:
\begin{equation}
    \begin{aligned}
    {\bf Y}_{\rm UL} \triangleq & \left[ \left( {\bf Y}_{\rm UL}^{(1)} \right)^T, \left( {\bf Y}_{\rm UL}^{(2)} \right)^T, \cdots, \left( {\bf Y}_{\rm UL}^{(N)} \right)^T \right]^T \\
    = &{\bf \Phi} {\bf H}_{\rm UL} + \tilde{\bf V}_{\rm UL},\\
    \label{eq:ul_full}
    \end{aligned}
\end{equation}
where
\begin{align}
{\bf \Phi} \triangleq &\left[ \left({\bf \Phi}^{(1)}\right)^T, \left({\bf \Phi}^{(2)}\right)^T, \cdots, \left({\bf \Phi}^{(N)}\right)^T \right]^T, \label{eq:sens_mat}\\
\tilde{\bf V}_{\rm UL} = &\left[ \left({\bf \Phi}^{(1)} {\bf V}_{\rm UL}^{(1)}\right)^T, \cdots, \left({\bf \Phi}^{(N)} {\bf V}_{\rm UL}^{(N)}\right)^T \right]^T. \label{eq:equiv_noise}
\end{align}

Leveraging TDD channel reciprocity \cite{smith2004direct}, we define the channel matrix as ${\bf H} = {\bf H}_{\rm UL}={\bf H}_{\rm DL}^H\triangleq [{\bf h}_1, \dots, {\bf h}_K]$. 
From the UL observations, the BS designs the analog precoder ${\bf F}_{\rm RF}\in\mathbb{C}^{M\times N_{\rm RF}}$ and the digital precoder ${\bf F}_{\rm BB}\in\mathbb{C}^{N_{\rm RF}\times K}$. 
The DL signal received by user $k$ is expressed as
\begin{align}
    y_{k, {\rm DL}} &= {\bf h}_k^H {\bf F}_{\rm RF} {\bf F}_{\rm BB} {\bf s} + v_{k, {\rm DL}} \nonumber \\
    &\triangleq \underbrace{{\bf h}_k^H {\bf F}_{\rm RF} {\bf f}_{{\rm BB}, k} s_k}_{\text{Signal}} + \underbrace{\sum\nolimits_{i\in\mathcal{K}, i \ne k} {\bf h}_k^H {\bf F}_{\rm RF} {\bf f}_{{\rm BB}, i} s_i}_{\text{Interference}} + v_{k, {\rm DL}},
    \label{eq:dl_user}
\end{align}
where ${\bf s} = [s_1, \dots, s_K]^T$ denotes the transmit symbol vector satisfying $\mathbb{E}[{\bf s} {\bf s}^H] = {\bf I}_K$. The total transmit power is constrained by $\| {\bf F}_{\rm RF} {\bf F}_{\rm BB} \|_F^2 \leq P_t$, and $v_{k, {\rm DL}} \sim \mathcal{CN}(0, \sigma_n^2)$ represents the AWGN at user $k$. Here, the digital precoder is decomposed as ${\bf F}_{\rm BB} = \left[{\bf f}_{{\rm BB}, 1}, \dots, {\bf f}_{{\rm BB}, K}\right]$. 
Stacking the signals for all $K$ users yields the global DL system model:
\begin{equation}
    {\bf y}_{\rm DL} = {\bf H}^H {\bf F}_{\rm RF} {\bf F}_{\rm BB} {\bf s} + {\bf v}_{\rm DL},
    \label{eq:dl_full}
\end{equation}
where ${\bf v}_{\rm DL} \sim \mathcal{CN}({\bf 0}, \sigma_n^2 {\bf I}_K)$ denotes the aggregate noise vector.

The signal-to-interference-plus-noise ratio (SINR) for user $k$ is given by
\begin{equation}
    \text{SINR}_k = \frac{ \left| {\bf h}_k^H {\bf F}_{\rm RF} {\bf f}_{{\rm BB}, k} \right|^2 }{ \sum_{i \neq k} \left| {\bf h}_k^H {\bf F}_{\rm RF} {\bf f}_{{\rm BB}, i} \right|^2 + \sigma_n^2 }.
    \label{eq:sinr}
\end{equation}
We maximize the system sum rate by jointly optimizing the analog and digital precoders:
\begin{subequations}\label{eq:sum_rate_prob}
    \begin{align}
    \underset{{\bf F}_{\rm RF}, {\bf F}_{\rm BB}}{\text{maximize}} \quad 
    & R = \sum_{k=1}^K \log_2 \left( 1 + \text{SINR}_k \right), \label{eq:opt_sum_rate} \\
    \text{subject to} \quad 
    & \left| \left[ {\bf F}_{\rm RF} \right]_{i,j} \right| = \frac{1}{\sqrt{M}}, \quad \forall i,j, \label{eq:cm_constraint} \\
    & \left\| {\bf F}_{\rm RF} {\bf F}_{\rm BB} \right\|_F^2 \leq P_t, \label{eq:power_constraint}
\end{align}
\end{subequations}
where the indices correspond to antennas $i=1, \cdots, M$ and RF chains $j=1, \cdots, N_{\rm RF}$. 
Problem \eqref{eq:sum_rate_prob} is nonconvex due to hardware-imposed CM constraint in \eqref{eq:cm_constraint}, and the multidimensional coupling between ${\bf F}_{\rm RF}$ and ${\bf F}_{\rm BB}$. 
While decoupled designs simplifies the problem but often suffer from performance degradation, conventional joint optimization via iterative algorithms entails high computational complexity, especially for large $M$. 
This motivates efficient strategies for XL-MIMO systems that avoid expensive computational overhead while capturing the coupled beamforming gain.

The boundary between the near- and far-field regions is commonly characterized by the Rayleigh distance $R_{\rm Rayleigh} = 2D^2/\lambda$ \cite{selvan2017fraunhofer}, where $D$ denotes the array aperture and $\lambda$ is the carrier wavelength. 
Within this region, conventional plane-wave models become inadequate as they fail to capture the non-negligible wavefront curvature. 
To address this limitation, we adopt a spherical-wave model to characterize the phase variation across the aperture. 
Accordingly, the channel vector for user $k$ is modeled as \cite{saleh1987statistical}
\begin{equation}
    {\bf h}_k = \sqrt{\frac{M}{L_k}} \sum_{\ell=1}^{L_k} \alpha_{\ell, k} e^{-j\frac{2\pi}{\lambda}r_{\ell, k}}{\bf b}(\theta_{\ell, k}, r_{\ell, k}), \label{eq:channel}
\end{equation}
where $L_k$ denotes the number of dominant paths, assumed to be identical across users without loss of generality, i.e., $L_k=L$, while $\alpha_{\ell, k}$, $\theta_{\ell, k}$, and $r_{\ell, k}$ denote the complex gain, the angle of arrival, and the distance from the array center to the user (or scatterer) of the $\ell$-th path, respectively. Notably, this formulation provides a unified framework that naturally reduces to the plane-wave model when the propagation distance is sufficiently large \cite{sherman1962properties}.

The near-field array response vector ${\bf b}(\theta, r) \in \mathbb{C}^{M\times1}$ for the ULA is defined as
\begin{equation}
    {\bf b}(\theta, r) = \frac{1}{\sqrt M}\left[e^{-j\frac{2\pi}{\lambda}\left(r^{(0)} - r \right)}, \cdots, e^{-j\frac{2\pi}{\lambda}\left(r^{(M-1)} - r \right)} \right]^T, \label{eq:array}
\end{equation}
where $r^{(m)} = \sqrt{r^2+\delta_m^2d^2-2r\delta_md\sin\theta}$ denotes the distance between the $m$-th antenna element and the target. Here, $d$ represents the antenna spacing, and $\delta_m = (2m-M+1)/2$ indexes the antenna positions relative to the array center, with $m=0, \dots, M-1$.

In multiuser near-field channels, jointly estimating the spherical-wave parameters $\left\{\alpha_{\ell,k},\theta_{\ell,k},r_{\ell,k}\right\}$ over an angle–range grid leads to a search space that grows with the total number of paths $KL$.
Moreover, spherical-wave dictionaries can be highly coherent along the range dimension and are sensitive to off-grid mismatch, which degrades estimation robustness.
Since channel estimation is only an intermediate step toward beamforming, the estimate-then-design pipeline can further suffer from error propagation, making estimation-based designs expensive for large arrays and sensitive to modeling mismatch.

\section{Near- and Far-Field Indirect E2E Beamforming with Perfect CSI}
In this section, we present the proposed E2E framework for indirect hybrid beamforming. The key ingredient is a variant-MMSE-based objective that eliminates the digital precoder in closed form, which stabilizes training and allows the analog stage to be learned under hybrid constraints. We describe the resulting complex-valued network architecture and summarize the inference procedure, where the predicted analog precoder is combined with the analytical digital solution to form the final hybrid beamformers.

\subsection{Decoupled Hybrid Beamforming via Sum-MSE Minimization}
We adopt the variant-MMSE criterion \cite{joham2005linear} as the primary training objective to guide the precoder optimization. 
Accordingly, the mean square error (MSE) is defined as the expected Euclidean distance between the transmitted signal vector $\bf s$ and the scaled received vector $\beta^{-1} \mathbf{y}_{\rm DL}$. 
Mathematically, the total sum-MSE is expressed as
\begin{equation}
    \begin{aligned}
        E &\triangleq \mathbb{E}_{{\bf s}, {\bf v}_{\rm DL}}\left\{\left\lVert {\bf s} - \beta^{-1}{\bf y}_{\rm DL} \right\rVert_2^2 \right\} \\
        &= {\rm tr}\big({\bf I}_K - \beta^{-1}{\bf F}_{\rm BB}^H{\bf F}_{\rm RF}^H{\bf H} - \beta^{-1}{\bf H}^H{\bf F}_{\rm RF}{\bf F}_{\rm BB} \\
        & \qquad\quad + \beta^{-2}{\bf H}^H{\bf F}_{\rm RF}{\bf F}_{\rm BB}{\bf F}_{\rm BB}^H{\bf F}_{\rm RF}^H{\bf H} + \beta^{-2}\sigma_n^2 {\bf I}_K \big),
    \end{aligned}\label{eq:MSEdef}
\end{equation}
where the scaling factor $\beta\in\mathbb{R}_{+}$ is introduced to align the amplitude of the effective received signal with the source ${\bf s}$, thereby facilitating optimization under the total power constraint \cite{joung2007regularized, stankovic2008generalized}. 
Minimizing this MSE effectively mitigates channel fading, MUI, and noise. 
Accordingly, we use the sum-MSE as a surrogate objective and formulate the joint hybrid beamforming problem as
\begin{subequations}
    \begin{align}
    \underset{{\bf F}_{\rm RF}, {\bf F}_{\rm BB}, \beta}{\rm minimize} \quad &E \label{eq:MSEmin_opt}\\
    {\rm s.t.} \quad &\left|[{\bf F}_{\rm RF}]_{i, j} \right| = \frac{1}{\sqrt M}, \quad \forall i,j,\label{eq:MSEmin_constant}\\
    & \left\lVert {\bf F}_{\rm RF}{\bf F}_{\rm BB} \right\rVert_F^2 \le P_t \label{eq:MSEmin_power}. 
    \end{align}
\end{subequations}

For a fixed ${\bf F}_{\rm RF}$, the optimization reduces to a convex problem with a single power constraint.
From the Karush--Kuhn--Tucker (KKT) stationarity condition with respect to ${\bf F}_{\rm BB}^*$, the digital precoder admits the closed form as
\begin{equation}
{\bf F}_{\rm BB} = \left({\bf F}_{\rm RF}^H{\bf H}{\bf H}^H{\bf F}_{\rm RF} + \lambda_{\rm Lag} {\bf F}_{\rm RF}^H{\bf F}_{\rm RF} \right)^{-1}{\bf F}_{\rm RF}^H{\bf H},
\end{equation}
where $\lambda_{\rm Lag}$ is the Lagrange multiplier associated with the power constraint.
Since the MSE objective is strictly decreasing with transmit power, the constraint must be active at the optimum, which implies $\lambda_{\rm Lag} = {K\sigma_n^2}/{P_t}$ for the MMSE criterion.
Factoring out the scaling term, the solution can be explicitly expressed as the product of $\beta$ and an unnormalized matrix $\tilde{\bf F}_{\rm BB}$:
\begin{equation}
    \begin{aligned}
        {\bf F}_{\rm BB} &= \beta\left({\bf F}_{\rm RF}^H{\bf HH}^H{\bf F}_{\rm RF} + \frac{K\sigma_n^2}{P_t}{\bf F}_{\rm RF}^H{\bf F}_{\rm RF} \right)^{-1}{\bf F}_{\rm RF}^H{\bf H} \\
        &= \beta\tilde{\bf F}_{\rm BB}. \label{eq:KKT}
    \end{aligned}
\end{equation}

Finally, substituting $\tilde{\bf F}_{\rm BB}$ into the active power constraint yields the closed-form scaling factor:
\begin{equation}
\beta = \sqrt{\frac{P_t}{{\rm tr} \left( {\bf F}_{\rm RF}\tilde{\bf F}_{\rm BB} \tilde{\bf F}_{\rm BB}^H  {\bf F}_{\rm RF}^H\right)}}. \label{eq:closed_beta}
\end{equation}

Substituting the closed-form expressions for $\beta$ and ${\bf F}_{\rm BB}$ back into \eqref{eq:MSEdef}, we derive the concentrated MSE objective function dependent solely on the analog precoder $\mathbf{F}_{\rm RF}$:
\begin{equation}
    \begin{aligned}
            E({\bf F}_{\rm RF}) = {\rm tr}\bigg( \Big({\bf I}_K + \frac{P_t}{K\sigma_n^2}{\bf H}^H{\bf F}_{\rm RF}&({\bf F}_{\rm RF}^H{\bf F}_{\rm RF})^{-1} \\
        &\quad\times {\bf F}_{\rm RF}^H{\bf H} \Big)^{-1} \bigg).
    \end{aligned}\label{eq:varMSE}
\end{equation}

In scenarios where the analog precoder ${\bf F}_{\rm RF}$ is approximately semi‑orthogonal, i.e., ${\bf F}_{\rm RF}^H{\bf F}_{\rm RF} \approx {\bf I}$, \eqref{eq:varMSE} reduces to the familiar Wiener or MMSE form. 
However, maintaining such orthogonality proves challenging in near-field environments, where the additional distance dimension often introduces significant spatial correlation among the analog beams. 
Therefore, we retain the exact objective function without approximation. 

The resulting single-variable optimization problem is formulated as
\begin{equation}
\underset{{\bf F}_{\rm RF}}{\rm min} \quad E({\bf F}_{\rm RF}) \qquad
{\rm s.t.} \quad \left|[{\bf F}_{\rm RF}]_{i, j} \right| = \frac{1}{\sqrt M}, \quad \forall i,j,\label{eq:opt_simp}
\end{equation}
which effectively decouples the design: for a fixed analog precoder ${\bf F}_{\rm RF}$, the digital precoder given in \eqref{eq:KKT} serves as the optimal solution to the convex sub-problem, leaving only the non-convex analog precoding problem to be solved. 
Compared with conventional disjoint two-stage schemes, this formulation facilitates a superior balance between signal enhancement and interference suppression, and the benefit becomes more pronounced as the number of users grows. This is particularly critical in near-field deployments, where users sharing similar angles but residing at different distances exhibit strong spatial correlation.
While iterative optimization algorithms can handle the CM constraints to find local optima, they typically require substantial iterations with computational complexity scaling as $\mathcal{O}(N_{\rm iter} M^2 K)$ \cite{gautam2023hybrid}, rendering them prohibitive for real-time ELAAs systems. 
To enable real-time inference without compromising performance, we learn a parametric mapping $f_{\Theta}:\ {\bf H}\mapsto{\bf F}_{\rm RF}$ that directly predicts the analog precoder from the CSI\footnote{The ``indirect'' approach necessitates explicit channel estimation as a prerequisite, distinguishing it from the ``direct'' approach in Section IV which maps pilots to beamformers without intermediate channel reconstruction.}.
Let $\Theta$ denote the network parameters.
We minimize the expected loss over the channel distribution:
\begin{equation}
\underset{\Theta}{\rm min} \quad \mathbb{E}_{\bf H}\left[ E({\bf F}_{\rm RF})\right], \qquad
{\rm s.t.} \quad
{\bf F}_{\rm RF} = f_{\Theta}({\bf H}),
\label{eq:opt_expect}
\end{equation}
where $f_{\Theta}(\cdot)$ represents the DNN mapping parameterized by $\Theta$.
In practice, the expectation is approximated via empirical averaging over a training dataset $\mathcal{T}$ of size $S$, i.e., $\mathcal{L}(\Theta) \approx \frac{1}{S}\sum_{s=1}^{S}E\left(f_{\Theta}\left({\bf H}^{(s)}\right)\right)$.
This formulation allows for efficient optimization via stochastic gradient descent (SGD).

\subsection{Proposed Complex-Valued E2E Network Architecture}
\begin{figure*}[t]
\centering
\includegraphics[width=6.6in]{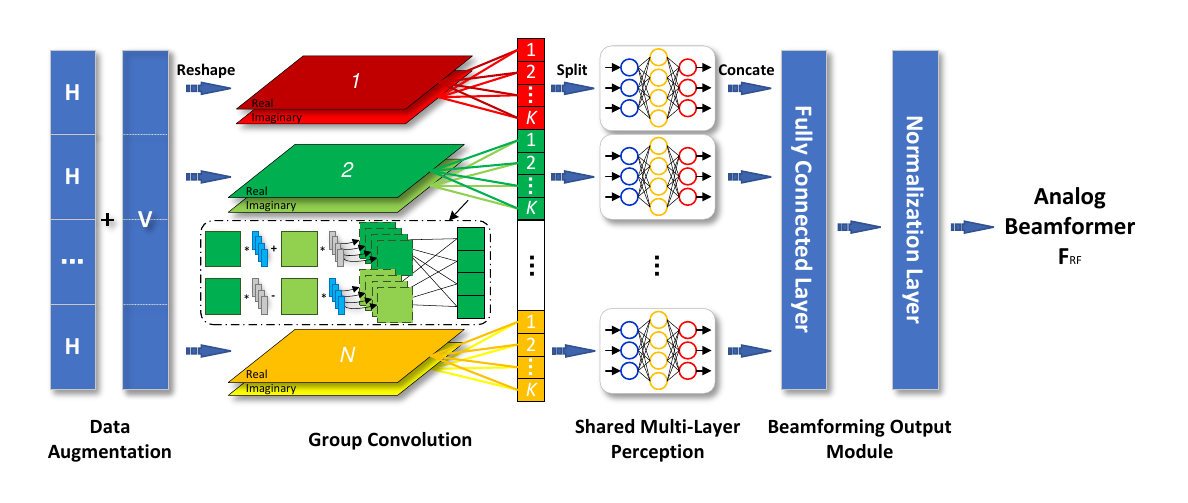}
\caption{Schematic of the proposed complex-valued E2E network. The model comprises: (i) a grouped complex-convolution sensing front-end that emulates the UL measurement process; (ii) a shared per-user complex MLP module for efficient feature extraction; and (iii) a merged output layer imposing CM normalization to generate ${\bf F}_{\rm RF}$.}
\label{fig:dnn}
\end{figure*}

As illustrated in Fig.~\ref{fig:dnn}, the proposed architecture takes the channel as input and consists of three modules: a grouped complex‑convolution module for virtual sensing, a shared per‑user complex MLP for latent feature extraction, and a beamforming output module culminating in a normalization layer to enforce the hardware constraints.

\subsubsection{Grouped Complex-Convolution Sensing Layer}

Designed to emulate the UL measurement process defined in \eqref{eq:ul_full}, this layer learns a bank of sensing matrices $\left\{{\bf \Phi}^{(n)}\right\}_{n=1}^N$ as part of the trainable network parameters $\Theta$. 
Structurally, each convolution group is assigned to a specific measurement slot to implement a distinct sensing matrix, which matches the physical time-domain beam switching mechanism. 
Compared to fully connected layers, the sparse connectivity inherent in grouped convolutions reduces the parameter space and provides implicit regularization. 
This improves robustness against noise, a critical advantage for large-scale antenna arrays. 
The extracted observation features are then forwarded to the downstream per‑user MLP for latent representation learning.

To accommodate the time-domain processing across $N$ measurement slots within a single coherence interval, the input channel matrix ${\bf H}^{(s)}$ is broadcasted along the temporal dimension. 
This yields an input tensor $\tilde{\bf H}^{(s)} \in \mathbb{C}^{N \times M \times K}$, where the $n$-th slice $\tilde{\bf H}^{(s)}_n = {\bf H}^{(s)}$ represents the static channel state. 
During the training phase, we superimpose an AWGN tensor $\tilde{\bf V}^{(s)} \in \mathbb{C}^{N \times M \times K}$ onto the inputs. 
The entries of $\tilde{\bf V}^{(s)}$ are drawn from an i.i.d. complex distribution $\mathcal{CN}(0,\sigma_n^2)$, resulting in the perturbed input $\tilde{\bf H}^{(s)}_{\rm noisy} = \tilde{\bf H}^{(s)} + \tilde{\bf V}^{(s)}$. 
This data augmentation strategy forces the network to learn robust features invariant to channel perturbations.
Note that this noise injection is exclusively applied during training to enhance generalization; it is disabled during the inference phase under the PCSI assumption. 

Unlike prior deep learning-based sensing architectures \cite{attiah2022deep, gao2022data} that restrict sensing weights to phase-only parameters under CM constraints, we implement bias-free complex-valued grouped convolutions.
By relaxing the CM constraint, this design significantly increases the sensing degrees of freedom, enabling the learning of optimal unconstrained linear combinations of antenna signals.
Furthermore, disabling bias terms strictly preserves the linearity of the signal model \eqref{eq:ul_slot}. 
Consequently, the learned kernels correspond exactly to the sequence of sensing matrices $\left\{{\bf \Phi}^{(n)}\right\}$, effectively capturing the temporal beam switching dynamics across measurement slots.

To implement this operation in standard deep learning frameworks, we adopt complex‑valued arithmetic proposed in \cite{Trabelsi2017deep}.
Specifically, the complex convolution is synthesized using two parallel real‑valued grouped convolution layers representing the real and imaginary kernels, respectively. 
These components are combined via complex algebra, i.e., $\left( \mathbf{W}_r * \mathbf{x}_r - \mathbf{W}_i * \mathbf{x}_i \right) + j\left( \mathbf{W}_r * \mathbf{x}_i + \mathbf{W}_i * \mathbf{x}_r \right)$, with bias terms strictly disabled. 
This ensures that the layer functions as a pure linear operator, maintaining compatibility with both indirect and direct beamforming schemes.

We configure the kernel size of the 2D grouped convolution as $(M,1)$ to span the entire antenna array. 
Consequently, for the $n$-th measurement slot, the $i$-th kernel represents the complex sensing vector 
$\bm{\phi}_{n,i} \in \mathbb{C}^{M}$ ($i=1,\dots,K$), corresponding to the $i$‑th row of the sensing matrix $\mathbf{\Phi}^{(n)}$. 
When applied to the $n$-th slice of the noisy input tensor $\tilde{\bf H}^{(s)}_{\rm noisy}$, the convolution operation performs a spatial projection equivalent to the inner product between the sensing vector and the channel.
Mathematically, the output for the $i$-th sensing channel is given by
\begin{equation} 
\begin{aligned} 
y_{n, i}    &= \bm{\phi}_{n,i} * \left[\tilde{\bf H}^{(s)}_{\rm noisy}\right]_{n,:,:} \\ 
            &= \left[ \bm{\phi}_{n,i}^T {\bf h}_1^{(s)}, \dots, \bm{\phi}_{n,i}^T {\bf h}_K^{(s)} \right] + \underbrace{\left[ \boldsymbol{\phi}_{n,i}^T \tilde{\bf v}_{n,1}^{(s)}, \dots, \boldsymbol{\phi}_{n,i}^T \tilde{\bf v}_{n,K}^{(s)} \right]}_{\text{Effective Noise}}, 
\end{aligned} 
\end{equation}
where $\tilde{\bf v}_{n,k}^{(s)} \triangleq [\tilde{\bf V}^{(s)}]_{n,:,k}$ denotes the injected noise vector for user $k$ at slot $n$. 
Aggregating the $K$ kernels constitutes the full sensing matrix ${\bf \Phi}^{(n)}$, where distinct groups learn independent matrices to mimic temporal beam switching. 
Through this design, the convolutional output rigorously aligns with the UL signal model $\mathbf{Y}_{\rm UL}$ in \eqref{eq:ul_full}, providing a physically interpretable input for the subsequent beamforming stages. 

To ensure structural consistency with the global UL signal model \eqref{eq:ul_full}, the outputs from all measurement slots are aggregated following the temporal stacking order. 
Specifically, let $\tilde{\bf Y}_n^{(s)}\in\mathbb{C}^{K\times K}$ denote the feature matrix generated by the grouped convolutions for the $n$-th slot of sample $s$. 
These per-slot features are vertically stacked to construct the comprehensive observation matrix: 
$
\tilde{\bf Y}^{(s)} \triangleq \left[(\tilde{\bf Y}_1^{(s)})^{T},\ldots,(\tilde{\bf Y}_N^{(s)})^{T} \right]^{T} \in \mathbb{C}^{NK\times K},
$
Finally, this aggregated matrix is sliced column‑wise to isolate specific user features, which serve as the input for the subsequent per-user multiplexing network.

In contrast to a generic fully connected sensing layer, which necessitates flattening the input into ${\rm vec}({\bf H})\in\mathbb{C}^{MK}$ and learning redundant correlations across independent users, the proposed grouped convolutional design employs $N\times K$ kernels of length $M$ to perform parallel per-user sensing.
This approach prevents over-parameterization by enforcing weight sharing across users, resulting in a compact model architecture that facilitates deployment on resource‑constrained hardware.

In summary, the grouped convolutional sensing layer functions as a learnable linear observation operator $f_{\bf \Phi}: {\bf H}\mapsto \tilde{\bf Y}$, parameterized by the matrix set $\{{\bf \Phi}^{(n)}\}_{n=1}^N$. 
This design effectively prevents over-parameterization and extracts noise-robust features, while strictly adhering to the structure of the physical model.

\subsubsection{Shared per‑user multiplexing module}
This module utilizes a shared complex‑valued MLP to independently extract latent features for each user. 
Specifically, for a given sample $s$, the input corresponding to user $k$ is extracted as the column vector $\tilde{\bf y}^{(s)}_{k} \triangleq [\tilde{\bf Y}^{(s)}]_{:,k} \in \mathbb{C}^{NK}$, derived from the aggregated observation matrix. 
By enforcing parameters sharing across users, the module achieves scalability regarding the user count $K$ and makes the module user-agnostic.
This design effectively reduces model complexity while enabling the learning of generalized and discriminative channel representations.

\begin{figure}[!t]
\centering
\includegraphics[width=0.45\textwidth]{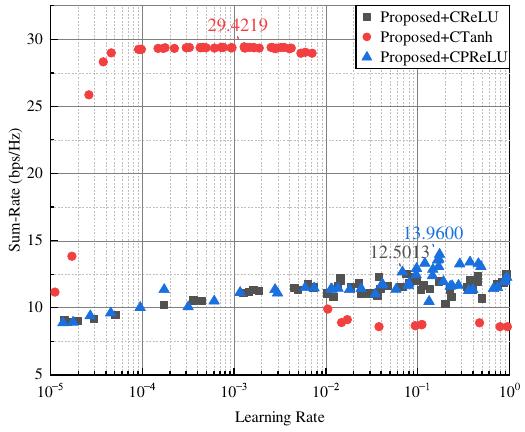}
\caption{Test sum‑rate with different complex activation functions under varying learning rates.}
\label{fig:actfunccomp}
\end{figure}

The MLP is composed of $P$ cascaded blocks, each comprising a complex-valued linear layer, a complex batch normalization (BN) layer, and a complex Tanh activation. 
Let ${\bf W}_p\in\mathbb{C}^{D_p\times D_{p-1}}$ and ${\bf b}_p\in\mathbb{C}^{D_p}$ denote the weight matrix and bias vector for the $p$‑th layer, respectively. 
Initializing the input as ${\bf x}_0=\tilde{\bf y}^{(s)}_{k}$, the operation of the $p$‑th block is formulated as
\begin{equation} 
\begin{aligned} 
\hat{\bf x}_p &= {\rm BN}_p\left({\bf W}_p{\bf x}_{p-1} + {\bf b}_p\right), \\ 
{\bf x}_p &= \mathbb{C}{\rm Tanh}\left(\hat{\bf x}_p\right), \qquad p=1,\ldots,P, 
\end{aligned} \label{eq:map_mlp} 
\end{equation}
where the complex Tanh function is applied element‑wise to both the real and imaginary components, defined as
\begin{equation}
\mathbb{C}{\rm Tanh}(x)\,\triangleq\,{\rm Tanh}(\mathfrak{R}(x))\;+\;j\,{\rm Tanh}(\mathfrak{I}(x)),
\label{eq:ctanh}
\end{equation}
with ${\rm Tanh}(t)=(e^{t}-e^{-t})/(e^{t}+e^{-t})$. 
As demonstrated in Fig.~\ref{fig:actfunccomp}, the proposed $\mathbb{C}$Tanh activation yields significantly higher sum rates compared to $\mathbb{C}$ReLU and $\mathbb{C}$PReLU. 
By preserving sign symmetry and bounding the output amplitude, $\mathbb{C}$Tanh effectively mitigates gradient explosion and vanishing issues in noisy regimes, which improves learning stability and convergence for complex-valued signals.

Complex BN \cite{Trabelsi2017deep} standardizes each complex feature by treating its real and imaginary components as a bivariate vector to perform whitening, followed by affine transformation:
\begin{equation}
{\rm BN}_p({\bf x}) \triangleq {\bf \gamma}_p\,\tilde{\bf x} + {\bf \beta}_p,\qquad
\tilde{\bf x} = {\bf C}^{-\frac{1}{2}}\!\big({\bf x}-\mathbb{E}[{\bf x}]\big),
\label{eq:complex_bn}
\end{equation}
where ${\bf C}$ denotes the $2 \times 2$ covariance matrix derived from the real--imaginary vector $[\mathfrak{R}({\bf x}),\mathfrak{I}({\bf x})]^T$. The terms ${\bf \gamma}_p\in\mathbb{R}^{2\times 2}$ and ${\bf \beta}_p\in\mathbb{C}$ represent the learnable scaling matrix and bias parameter, respectively. 

Cascading these $P$ blocks establishes the per‑user mapping $f_{\zeta}:\tilde{\bf y}_k \mapsto {\bf x}_P$ for each user $k$. This architecture facilitates stable feature extraction and dimensionality expansion\footnote{In the direct beamforming mode, the network takes low-dimensional pilot observations as input and outputs beamforming-related features for the downstream module. } while functioning as a user‑agnostic analog beamforming preprocessor.

\subsubsection{Merged Beamforming Output Network}
To effectively manage inter‑user interference, the merged beamforming output network aggregates the latent features from all users to generate the analog precoder. 
Let ${\bf x}_P^{(k,s)}\in\mathbb{C}^{D_P}$ denote the extracted feature vector for user $k$ corresponding to sample $s$. 
The feature vectors of the $K$ users are concatenated in a fixed sequence to construct the global feature vector:
\begin{equation}
    {\bf z}^{(s)} \triangleq \left[\left({\bf x}_P^{(1,s)}\right)^{T}, \ldots, \left({\bf x}_P^{(K,s)}\right)^{T} \right]^{T} \in \mathbb{C}^{K D_P}.
\end{equation}

Subsequently, a complex-valued linear layer, parameterized by the set $\{{\bf W}_{\rm final},{\bf b}_{\rm final}\}\subset\Theta$, maps ${\bf z}^{(s)}$ to the vectorized representation of the unnormalized analog precoder:
\begin{equation}
\operatorname{vec}\left(\tilde{\bf F}_{\rm RF}^{(s)}\right) = {\bf W}_{\rm final} {\bf z}^{(s)} + {\bf b}_{\rm final}\in \mathbb{C}^{M N_{\rm RF}},
\label{eq:unnorm_frf}
\end{equation}
where ${\bf W}_{\rm final}\in\mathbb{C}^{MN_{\rm RF}\times K D_P}$ and ${\bf b}_{\rm final}\in\mathbb{C}^{MN_{\rm RF}}$ constitute the weight matrix and bias vector, respectively. 

To satisfy the hardware-imposed CM constraint, we apply a normalization function on each element while ensuring unit-norm power constraints for the precoding vectors:
\begin{equation}
{\rm N}_{\varepsilon}({\bf X}) \triangleq\frac{{\bf X}\oslash(|{\bf X}|+\varepsilon)}{\sqrt{M}}, \qquad \varepsilon>0,
\label{eq:cm_norm}
\end{equation}
where $\oslash$ denotes the Hadamard division, and $|\bf X|$ represents the element-wise magnitude matrix. A regularization term $\varepsilon$ is introduced to enhance numerical stability by preventing division-by-zero errors during training. 
Consequently, the final analog precoder is obtained via
\begin{equation}
{\bf F}_{\rm RF}^{(s)} = {\rm N}_{\varepsilon}\left(\tilde{\bf F}_{\rm RF}^{(s)}\right)
\triangleq f_{\rm final}\left({\bf z}^{(s)};{\bf W}_{\rm final},{\bf b}_{\rm final}\right),
\label{eq:map_outlayer}
\end{equation}
which retains the phase information of the unnormalized matrix $\tilde{\bf F}_{\rm RF}^{(s)}$ while normalizing the magnitude of each entry to $1/\sqrt{M}$. This joint mapping mechanism enables the explicit coordination of spatial beams across RF chains. This design is crucial for mitigating inter‑user interference prior to the subsequent digital precoding stage.

\subsection{Deep Learning-Based Near- and Far-Field Indirect Multiuser Hybrid Beamforming}
Leveraging the decoupled optimization framework, the optimal digital precoder is substituted via its closed-form KKT solutions, which leads to a sum‑MSE objective that depends on ${\bf F}_{\rm RF}$. 
Consistent with the problem formulation in \eqref{eq:opt_expect}, the proposed complex-valued network is trained to minimize the following loss function:
\begin{equation}
\begin{aligned}
\mathcal{L}({\bf F}_{\rm RF}) = {\rm tr}\bigg(\Big({\bf I}_K + \frac{P_t}{K\sigma_n^2}\,{\bf H}^H{\bf F}_{\rm RF}(&{\bf F}_{\rm RF}^H {\bf F}_{\rm RF})^{-1} \\ 
&\times{\bf F}_{\rm RF}^H{\bf H}\Big)^{-1}\bigg), \label{eq:lossfn}
\end{aligned}
\end{equation}
which is differentiable with respect to ${\bf F}_{\rm RF}$, provided that the system satisfies $N_{\rm RF}\ge K$ and that ${\bf F}_{\rm RF}$ maintains full column rank. 

Under the assumption of PCSI availability during both training and inference phases, the loss function defined in \eqref{eq:lossfn} is evaluated directly using the channel matrix ${\bf H}$. 
Since ${\bf H}$ serves both the input to the neural network, and the parameter for the loss function, the training does not require supervised label. 
For a mini‑batch of $S$ channel realizations $\{ {\bf H}^{(s)} \}_{s=1}^{S}$, the empirical optimization problem is formulated as
\begin{equation}
\min_{\Theta}\quad\frac{1}{S}\sum_{s=1}^{S}\mathcal{L}\left({\bf F}_{\rm RF}^{(s)}\right),\qquad
{\rm s.t.}\quad{\bf F}_{\rm RF}^{(s)}=f_{\Theta}\left({\bf H}^{(s)}\right),
\end{equation}
where $f_{\Theta}(\cdot)$ represents the E2E neural network mapping, parameterized by $\Theta$, which generates the analog precoder subject to the CM constraint. 

\begin{algorithm}[!t]
\caption{Deep Learning‑Based Near- and Far‑Field Indirect Multiuser Hybrid Beamforming (DL-IMHB)}
\label{alg:pcsi}
\begin{algorithmic}[1]
\REQUIRE Channel matrix ${\bf H}\in\mathbb{C}^{M\times K}$ via PCSI or estimation, noise power $\sigma_n^2$, number of users $K$, maximum transmit power $P_t$, and optimal network parameters $\Theta^\star$
\ENSURE Optimal analog precoder ${\bf F}_{\rm RF}^\star\in\mathbb{C}^{M\times N_{\rm RF}}$ and digital precoder ${\bf F}_{\rm BB}^\star\in\mathbb{C}^{N_{\rm RF}\times K}$
\STATE Initialize the neural network with the pre-trained parameters $\Theta^\star$.
\STATE Generate the CM‑normalized analog precoder via forward propagation: ${\bf F}_{\rm RF}^\star \leftarrow f_{\Theta^\star}({\bf H})$.
\STATE Compute the unnormalized digital precoder $\tilde{\bf F}_{\rm BB}$ via the closed-form solution \eqref{eq:KKT} given ${\bf H}$ and ${\bf F}_{\rm RF}^\star$.
\STATE Derive the power scaling factor $\beta$ according to \eqref{eq:closed_beta} to satisfy the total transmit power constraint.
\RETURN Final hybrid beamforming matrices: ${\bf F}_{\rm RF}^\star$ and ${\bf F}_{\rm BB}^\star \leftarrow \beta \tilde{\bf F}_{\rm BB}$.
\end{algorithmic}
\end{algorithm}

After training, the inference procedure for of the proposed deep learning–based near‑ and far‑field indirect hybrid beamforming (DL-IMHB) is summarized in Algorithm~\ref{alg:pcsi}. 
In this framework, the estimated channel matrix ${\bf H}$ serves as the explicit input for generating both the analog precoder ${\bf F}_{\rm RF}$ and the subsequent closed‑form digital precoder.
This dependency characterizes the indirect approach to hybrid beamformer design. 
When instantiated with PCSI, this indirect approach effectively establishes a theoretical performance upper bound for the corresponding measurement-driven direct beamforming schemes.
Although the same architecture supports both modes, $N$ has different meanings in the indirect and direct settings.
In the direct mode, $N$ dictates the over-the-air pilot overhead; conversely, in this indirect mode, $N$ controls the dimensionality of the virtual sensing front-end and is independent of the pilot budget.

As detailed above, the indirect strategy exploits explicit channel knowledge, utilizing ${\bf H}$ as the input to generate ${\bf F}_{\rm RF}$ while deriving the digital precoder analytically.
However, in direct beamforming scenarios without explicit CSI, the primary objective shifts to designing a robust architecture capable of adapting to rapid near‑ and far‑field channel fluctuations, thereby facilitating fast multiuser access using short pilot sequences even in noisy environments.
The proposed framework accommodates different sensing dimension $N$: a larger $N$ enhances the sensing degrees of freedom, yielding superior spectral efficiency when pilot resources are abundant, whereas a smaller $N$ prioritizes low-latency operation with minimal overhead.

\section{Robust Direct End-to-End Hybrid Beamforming with Short Pilots}
This section investigates the direct hybrid beamforming paradigm, designed for scenarios where instantaneous CSI is unavailable during inference.
Building upon the unified architecture established in Section~III, we deploy the proposed grouped complex-convolution sensing layer to enable implicit channel feature extraction.
Unlike the indirect approach that relies on explicit channel inputs, the direct paradigm relies on a distinct inference mechanism, where the kernels learned by the grouped convolution layer are extracted and mapped to configure the physical analog sensing matrices $\left\{{\bf \Phi}^{(n)}\right\}$.
This strategy allows the hardware front-end to physically capture pilot observations and directly feed the received signals into the subsequent inference network, thereby avoiding explicit estimation of the full-dimensional channel during inference.

\subsection{Problem Reformulation}
In contrast to the sum‑MSE optimization in Section~III, the direct beamforming paradigm assumes that the explicit channel matrix ${\bf H}$ is unavailable in the inference phase. 
The network therefore takes only the UL measurement matrix ${\bf Y}_{\rm UL}$, acquired via the sensing interface, as input. 
To integrate the physical sensing process into the learning framework, the initial sensing stage is parameterized as a learnable grouped-convolution layer that implements the measurement operation:
\begin{equation}
    \begin{aligned}
    &{\bf Y}_{\rm UL} = f_{\bf \Phi} \left({\bf H},\tilde{\bf V}_{\rm UL}\right),\\
    f_{\bf \Phi} \Big({\bf H}&, \tilde{\bf V}_{\rm UL}\Big) \equiv \mathcal{G}_{\bf \Phi}({\bf H}) + \tilde{\bf V}_{\rm UL},
    \end{aligned}
\end{equation}
where $\mathcal{G}_{\bf \Phi}(\cdot)$ denotes the linear projection induced by the grouped convolutions, consistent with \eqref{eq:ul_full}, and $\tilde{\bf V}_{\rm UL}$ is consistent with its definition in \eqref{eq:equiv_noise}. 
Learning ${\bf \Phi}$ amounts to optimizing the physical measurement design jointly with the subsequent beamforming network.

The downstream beamforming network consists of a shared per‑user MLP module $f_{\zeta}(\cdot)$, followed by a merged output head $f_{\rm final}(\cdot)$ that incorporates the CM normalization layer. 
The parameter set $\Theta_{\rm NCSI}\triangleq\left\{{\bf \Phi},\zeta,{\bf W}_{\rm final},{\bf b}_{\rm final}\right\}$ is jointly optimized by minimizing the expected sum-MSE objective in Section~III, where the analog precoder is obtained from the UL measurements:
\begin{equation}
\begin{aligned}
\underset{\Theta_{\rm NCSI}}{\rm minimize}\quad
& \mathbb{E}_{{\bf H},\tilde{\bf V}_{\rm UL}}\left[\mathcal{L}\left({\bf F}_{\rm RF};{\bf H}\right)\right], \\
{\rm s.t.}\quad
& {\bf Y}_{\rm UL}= f_{\bf \Phi}\left({\bf H},\tilde{\bf V}_{\rm UL}\right),\\
& {\bf F}_{\rm RF}= f_{\rm final}\left(f_{\zeta}({\bf Y}_{\rm UL})\right),
\end{aligned}\label{eq:ncsi_opt}
\end{equation}
where $\mathcal{L}({\bf F}_{\rm RF};{\bf H})$ corresponds to the sum‑MSE metric defined in \eqref{eq:lossfn}.
Crucially, although the channel realization ${\bf H}$ is used to evaluate the loss during training, it is not fed into the network and serves only as supervision.  
During the inference stage, the system operates exclusively on the acquired measurements ${\bf Y}_{\rm UL}$ to predict the analog precoder ${\bf F}_{\rm RF}$, ensuring blind operation without instantaneous CSI. 

Two implementation details clarify the sensing interface.
First, the sensing matrix ${\bf \Phi}$ is not subject to the analog CM constraint, since it represents the composite effect of the RF analog sensing and digital baseband processing. 
This increases the sensing degrees of freedom while keeping the downstream analog precoder CM-compliant.
Second, the grouped‑convolution parametrization matches the across‑antenna linear projections. Bias terms are omitted during training to preserve the linear sensing model in \eqref{eq:ul_full}. 

\subsection{Effective Channel Estimation and Digital Beamforming Protocol}
\begin{figure}[!t]
\centering
\includegraphics[width=3.5in]{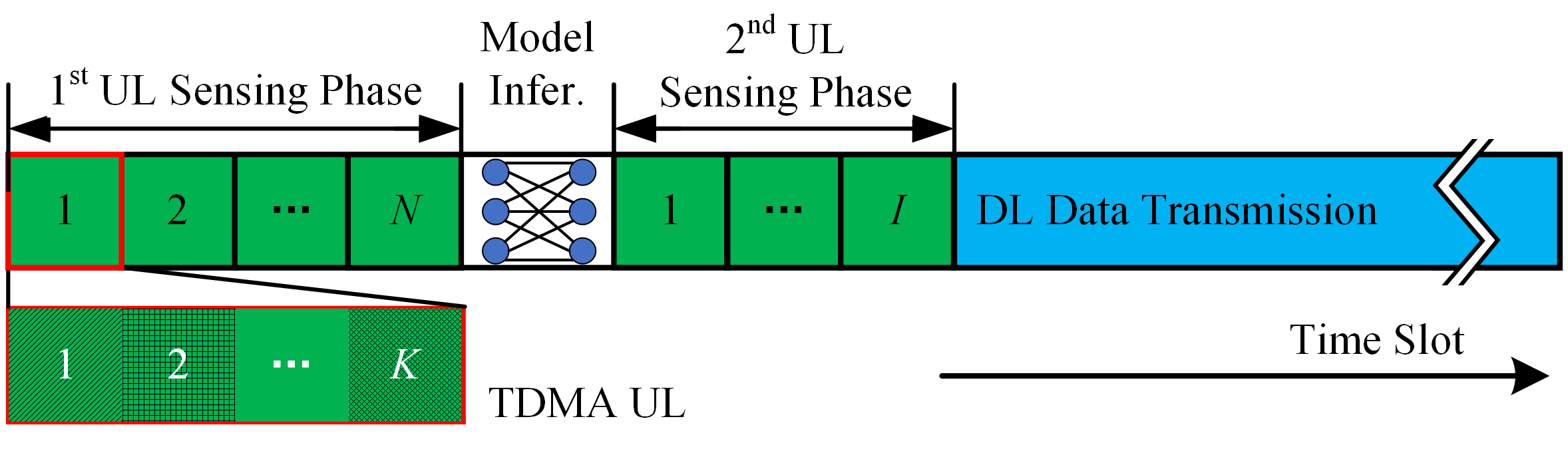}
\caption{Timeline of signaling protocol, comprising TDMA UL sensing for measurement acquisition, neural network inference for analog precoder generation, effective channel estimation via pilot repetition, and subsequent downlink data transmission.}
\label{fig:signaling}
\end{figure}

In the direct beamforming paradigm, a fundamental challenge is that the unnormalized digital precoder $\tilde{\bf F}_{\rm BB}$ cannot be derived from the full channel ${\bf H}$ due to the lack of explicit CSI.
To address this, we employ a multi-stage signaling protocol, depicted in Fig.~\ref{fig:signaling}, which facilitates the estimation of the low-dimensional effective channel ${\bf H}_{\rm eq} \triangleq {\bf H}^H {\bf F}_{\rm RF} \in \mathbb{C}^{K\times N_{\rm RF}}$ to support the subsequent digital precoder construction. Here, $N$ is the number of UL sensing slots used for analog inference, and $I$ is the number of repetition blocks used to estimate the effective channel ${\bf H}_{\rm eq}$.

Upon generation of the CM‑normalized analog precoder ${\bf F}_{\rm RF}^\star$ by the DNN as described in Section IV-B, this matrix is fixed and deployed as the UL analog combiner.
Utilizing a transmission protocol with $I$ repetition blocks such as TDMA, the baseband observation matrix for the $i$‑th block is formulated as
\begin{equation}
\bar{\bf Y}_{\rm UL}^{(i)} = \left({\bf F}_{\rm RF}^\star\right)^{H}\left({\bf H} + {\bf V}_{\rm UL}^{(i)}\right) \in \mathbb{C}^{N_{\rm RF}\times K},\quad i=1,\ldots,I,
\label{eq:ncsi_ul}
\end{equation}
where the $k$-th column of $\bar{\bf Y}_{\rm UL}^{(i)}$ represents the received pilot signal corresponding to user $k$ isolated in the time domain.
Averaging these observations across the $I$ repetitions yields the estimated effective channel:
\begin{equation}
\hat{\bf H}_{\rm eq} = \left(\frac{1}{I}\sum_{i=1}^{I}\bar{\bf Y}_{\rm UL}^{(i)}\right)^{H} \;\in\; \mathbb{C}^{K\times N_{\rm RF}}.
\label{eq:ncsi_heq}
\end{equation}
Averaging over $I$ repetitions reduces the estimation noise and provides a practical accuracy-latency trade-off.

Based on the estimated effective channel $\hat{\bf H}_{\rm eq}$, the closed-form solution for the digital precoder is derived consistent with the variant-MMSE criterion:
\begin{equation}
\hat{\tilde{\mathbf{F}}}_{\rm BB} = \left(\hat{\bf H}_{\rm eq}^H \hat{\bf H}_{\rm eq} + \frac{K\sigma_n^2}{P_t} {\bf F}_{\rm RF}^H{\bf F}_{\rm RF} \right)^{-1}\hat{\bf H}_{\rm eq}^H.
\label{eq:ncsi_fbb}
\end{equation}
For numerical implementation, the inversion of the Gram matrix is regularized as $({\bf A}^H{\bf A}+\varepsilon{\bf I})^{-1}$ with a small damping factor $\varepsilon>0$. The system is solved via Cholesky factorization with forward and backward substitution to avoid explicit inversion. The final digital precoder is then obtained by power normalization via $\hat{\beta}$ as in \eqref{eq:closed_beta}.

\begin{algorithm}[!t]
\caption{Deep Learning‑Based Near‑ and Far-Field Direct Multiuser Hybrid Beamforming (DL-DMHB)}
\label{alg:ncsi}
\begin{algorithmic}[1]
\REQUIRE Noise power $\sigma_n^2$, number of users $K$, maximum transmit power $P_t$, sensing pilot length $N$, number of repetition blocks $I$, and trained parameter set $\Theta^\star = \{{\bf \Phi}^\star, \zeta^\star, {\bf W}_{\rm final}^\star, {\bf b}_{\rm final}^\star \}$
\ENSURE Analog precoder ${\bf F}_{\rm RF}^\star \in \mathbb{C}^{M\times N_{\rm RF}}$ and estimated digital precoder $\hat{\bf F}_{\rm BB}^\star \in \mathbb{C}^{N_{\rm RF}\times K}$
\STATE Initialize the sensing interface with ${\bf \Phi}^\star$ and the inference network with the remaining parameters ${\Theta}^\star\setminus{\bf \Phi}^\star$.
\FOR{$n=1$ to $N$}
    \STATE Configure the sensing combiner weights according to the sub-matrix ${\bf \Phi}^{(n)}$
    \STATE Receive UL TDMA pilot signals from all $K$ users
    \STATE Acquire the measurement snapshot ${\bf Y}_{\rm UL}^{(n)}$ via \eqref{eq:ul_slot}
\ENDFOR
\STATE Construct the aggregated measurement matrix ${\bf Y}_{\rm UL}$ as defined in \eqref{eq:ul_full}
\STATE Generate the CM‑normalized analog precoder via forward propagation: ${\bf F}_{\rm RF}^\star \leftarrow f_{\rm final} \big(f_{\zeta}({\bf Y}_{\rm UL})\big)$
\STATE Deploy ${\bf F}_{\rm RF}^\star$ as the fixed UL analog combiner
\FOR{$i=1$ to $I$}
    \STATE Record the baseband observation $\bar{\bf Y}_{\rm UL}^{(i)}$ following RF combining as in \eqref{eq:ncsi_ul}
\ENDFOR
\STATE Calculate the effective channel estimation $\hat{\bf H}_{\rm eq}$ via averaging \eqref{eq:ncsi_heq}
\STATE Derive the unnormalized digital precoder $\hat{\tilde{\bf F}}_{\rm BB}$ using \eqref{eq:ncsi_fbb} via Tikhonov regularization and Cholesky‑based linear solving
\RETURN the final hybrid beamformers: ${\bf F}_{\rm RF}^\star$ and $\hat{\bf F}_{\rm BB}^\star \leftarrow \hat{\beta}\,\hat{\tilde{\bf F}}_{\rm BB}$
\end{algorithmic}
\end{algorithm}

The comprehensive operational workflow, integrating the sensing front-end, DNN inference, and the aforementioned digital beamforming stages, is summarized in Algorithm~\ref{alg:ncsi}.
The number of pilot repetitions $I$ in the protocol can be fixed for benchmarking or determined adaptively based on a convergence threshold $\epsilon$. Alternatively, the digital precoder allows for iterative refinement using WMMSE algorithms initialized by \eqref{eq:ncsi_fbb}.

\section{Numerical Results and Discussion}
This section presents numerical results for the proposed PCSI-driven DL-IMHB and measurement-driven DL-DMHB under representative near- and far-field channel conditions. Their performance is compared with existing baseline schemes as well as with fully digital or PCSI-aided reference solutions.

\subsection{System Setup}
An XL-MIMO system operating in TDD mode at a carrier frequency of 100~GHz is considered, where the base station employs a 128-antenna ULA. Unless otherwise specified, the number of RF chains is set to $N_{\rm RF}=K=4$.
The uplink channel model follows \eqref{eq:channel}, with half-wavelength antenna spacing. For each channel realization, the complex path gains are drawn from $\mathcal{CN}(0,1)$, the angle cosines are uniformly distributed over $[-1,1]$, and the path distances are uniformly distributed in the range from 5~m to $r_{\max}$, where $r_{\max}=80$~m unless stated otherwise.
Under these settings, the corresponding Rayleigh distance is approximately 24.19~m, so that both near-field and far-field propagation regimes are included.
The total transmit power and the reference receive power are normalized to unity, which yields identical uplink and DL SNR values given by ${\rm SNR}_{\rm UL} = {\rm SNR}_{\rm DL} = 10\log_{10}\left(1/\sigma_n^2\right)$.

Independent datasets are constructed for different system configurations specified by the user count $K$, the number of dominant paths $L$, and the maximum path distance $r_{\max}$.
Each dataset contains 20,000 independent multiuser channel realizations, which are divided into training, validation, and test sets with a ratio of 70\%, 15\%, and 15\%, respectively.

\begin{table}[!t]
\caption{Simulation Experiment Hyperparameter Configuration}
\label{tab:simhyperset}
\centering
\begin{tabular}{|c|c|}
\hline
\textbf{Hyperparameter} & \textbf{Value} \\ \hline
Initial Learning Rate & $5\times 10^{-4}$ \\ \hline
Batch Size & 1024 \\ \hline
Optimizer & Adam \\ \hline
Learning Rate Scheduler & ReduceLROnPlateau \\ \hline
Early Stopping Threshold & $1\times10^{-4}$ \\ \hline
Scheduler Threshold & $1\times10^{-4}$ \\ \hline
Early Stopping Patience & $3\times10^2$ \\ \hline
Scheduler Patience & $2\times10^2$ \\ \hline
Scheduler Scaling Factor & $0.5$ \\ \hline
\end{tabular}
\end{table}

The deep learning algorithms are implemented in PyTorch using an NVIDIA GeForce RTX 4070 GPU. 
Early stopping monitors the average validation sum rate.
Hyperparameter optimization is conducted using the Optuna framework~\cite{optuna_2019}, with specific configuration settings detailed in Table~\ref{tab:simhyperset}. 
The shared multiplexing module is instantiated with $P=3$ cascaded blocks, featuring sequential feature dimensions of $D_1 = 512$, $D_2 = 256$, and $D_3 = 128$.

\subsection{Baselines and Computational Complexity}
The baselines are grouped into the PCSI-based indirect regime in Algorithm~\ref{alg:pcsi} and the measurement-driven direct regime in Algorithm~\ref{alg:ncsi}.
Performance metrics include sum rate defined in \eqref{eq:opt_sum_rate} and the variant-MMSE-based objective derived in \eqref{eq:varMSE}; specifically, for Fig.~\ref{fig:pcsi_core_1}(b), the latter is visualized on a logarithmic scale as $10\log_{10}(E({\bf F}_{\rm RF}))$ dB to visualize the optimization objective on a logarithmic scale. 
Furthermore, to assess scalability, Fig.~\ref{fig:pcsi_core_2} and Fig.~\ref{fig:perf_pcsi_numusers} utilize the normalized per-user average sum rate, calculated as $R/K$.

\begin{table*}[t]
\begin{center}
\caption{Computational Complexity, Runtime, and Configuration Comparison.}
\label{tab:complexity}
\scriptsize
\setlength{\tabcolsep}{4pt}
\renewcommand{\arraystretch}{1.5}
\begin{tabular}{| c | l | c | c | c |}
\hline
\textbf{Setting} & \textbf{Algorithm} & \textbf{Computational Complexity} & \textbf{Avg. Runtime (ms)} & \textbf{Configuration} \\
\hline\hline

\multirow{4}{*}{\textbf{PCSI}}
& DL-IMHB 
& $\mathcal{O}((N + D_P)K^2 M + K\sum_{p=1}^P D_{p-1}D_p)$ 
& 0.68 (GPU) / 2.89 (CPU) 
& $N=8$ \\ \cline{2-5}

& SU-DNN
& $\mathcal{O}(NK^2 M + K\sum_{p=1}^P D_{p-1}D_p)$ 
& 0.44 (GPU) / 1.72 (CPU) 
& $N=8$ \\  \cline{2-5}

& TH-HMP 
& $\mathcal{O}(N_{\rm iter}KM^2)$ 
& 5.82 (CPU) 
& $N_{\rm iter}=10$ \\ \cline{2-5}

& LDMA 
& $\mathcal{O}(SKM+K^2 M)$ 
& 26.57 (CPU) 
& $\beta = 0.4$ \\ \cline{2-5}
\hline\hline

\multirow{4}{*}{\textbf{NCSI}}
& DL-DMHB
& $\mathcal{O}(D_P K^2 M + K\sum_{p=1}^P D_{p-1}D_p)$ 
& 0.53 (GPU) / 2.31 (CPU) 
& $N=12, I=4$ \\ \cline{2-5}

& SU-DNN
& $\mathcal{O}(K\sum_{p=1}^P D_{p-1}D_p)$ 
& 0.42 (GPU) / 1.31 (CPU) 
& $N=12, I=4$ \\ \cline{2-5}

& P-SOMP 
& $\mathcal{O}(LSNKM)$ 
& 13.45 (CPU) 
& $N=24$, $\beta = 1$ \\ \cline{2-5}

& P-SIGW 
& $\mathcal{O}(LSNKM + N_{\rm iter}L^2 N K M)$ 
& 464.83 (CPU) 
& $N=24$, $\beta=1$, $N_{\rm iter}=20$ \\
\hline
\end{tabular}
\end{center}
\end{table*}

\begin{table}[!t]
\centering
\caption{Lifecycle Computational Complexity Analysis.}
\label{tab:lifecycle}
\begin{tabular}{l c| c|| c| c}
\toprule
\multirow{2}{*}{\textbf{Metric}} & 
\textbf{Proposed} & 
\textbf{Proposed} & 
\textbf{SU-DNN} & 
\textbf{SU-DNN} \\
 & 
\textbf{DL-DMHB} & 
\textbf{DL-IMHB} & 
\textbf{w/o CSI} & 
\textbf{w/ CSI} \\
\midrule 
Learnable Params. & 0.92 M & 0.90 M & 0.39 M & 0.37 M \\
Storage Footprint & $\sim$3.7 MB & $\sim$3.6 MB & $\sim$1.6 MB & $\sim$1.5 MB \\
\midrule
Training Time & $\sim$24.4 min & $\sim$29.1 min & $\sim$11.5 min & $\sim$9.8 min \\
\midrule 
Theoretical FLOPs & 8.21 M & 8.03 M & 6.11 M & 5.82 M \\
\bottomrule
\end{tabular}
\vspace{1ex}
\flushleft
\end{table}

We analyze inference computational complexity assuming $N_{\rm RF}=K$ and $M \gg \{K,N,L,S\}$, with lower-order terms omitted for brevity. 
The grouped-convolution sensing module contributes an $\mathcal{O}(NK^2M)$ term in the indirect mode, while it is not executed in the direct mode where measurements are acquired over the air.
For P-SOMP and P-SIGW, we exclude offline costs like dictionary whitening and codebook construction. 
Table~\ref{tab:complexity} summarizes the dominant-order inference complexities and the corresponding average runtimes for each algorithm under their specific configurations. 
Table~\ref{tab:lifecycle} further details the model size, measured training time, and theoretical floating-point operations (FLOPs), utilizing configurations consistent with Table~\ref{tab:complexity}. 

\subsubsection{Indirect beamforming baselines}
The following baselines assume explicit CSI availability: 
\begin{itemize}
\item \textbf{SU-DNN \cite{attiah2022deep} w/ CSI}: DL-based single-carrier hybrid beamforming framework. The analog sensing layer is configured to perform $N$ virtual measurements. Subsequently, the system utilizes PCSI to construct the equivalent baseband channel, deriving the digital precoder via ZF. 
\item \textbf{LDMA \cite{wu2023multiple}}: Exhaustive angle--distance codebook search, with the minimum distance constraint set to $\rho_{\min}=5$ m. The codebook granularity---specifically the number of distance rings $S$---is governed by the threshold parameter $\beta$.
\item \textbf{TH-HMP \cite{gautam2023hybrid}}: Iterative SVD-based algorithm designed to minimize the sum-MSE objective. 
\item \textbf{Fully Digital}: Unconstrained fully-digital beamforming solution, serving as the theoretical performance benchmark for hybrid architectures.
\item \textbf{DL-IMHB}: Proposed indirect multiuser hybrid beamforming framework as shown in Algorithm~\ref{alg:pcsi}, utilizing the deep complex network architecture detailed in Section~III-B. 
\end{itemize}

\subsubsection{Direct beamforming baselines}
The following baselines derive beamformers directly from received pilot signals:
\begin{itemize}
\item \textbf{SU-DNN \cite{attiah2022deep} w/o CSI}: In the inference phase, the pre-trained sensing layer is frozen and deployed as the analog combiner to acquire $N$ measurements. Subsequently, $I$ pilot repetition blocks are utilized to estimate the equivalent baseband channel via linear MMSE (LMMSE) estimation, followed by ZF digital precoding.
\item \textbf{P-SOMP \cite{cui2022channel}}: Orthogonal matching pursuit algorithm adapted for polar-domain sparsity. It utilizes the same codebook configuration as LDMA, with $\beta$ in $\{0.6,1\}$ and the target sparsity level $\hat L=2L$ to capture multipath components.
\item \textbf{P-SIGW \cite{cui2022channel}}: Gradient-based refinement scheme initialized by P-SOMP to fine-tune angle and distance estimates. The algorithm executes for $N_{\rm iter}=20$ iterations using pre-whitened input signals, with hyperparameter optimization performed via Optuna.
\item \textbf{DL-DMHB}: Proposed direct hybrid beamforming framework as outlined in Algorithm~\ref{alg:ncsi}, employing the architecture described in Section~III-B. 
\end{itemize}

\subsection{Indirect Hybrid Beamforming Performance}
\begin{figure}[!t]
\centering
\subfloat[Achievable sum rate versus algorithm-specific complexity parameters]{\includegraphics[width=0.4\textwidth]{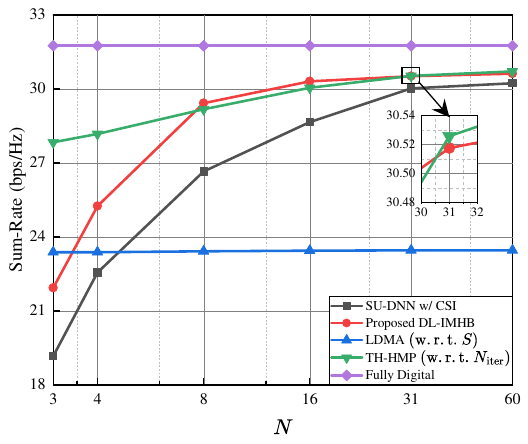}} \\
\subfloat[Sum-MSE versus algorithm-specific complexity parameters]{\includegraphics[width=0.4\textwidth]{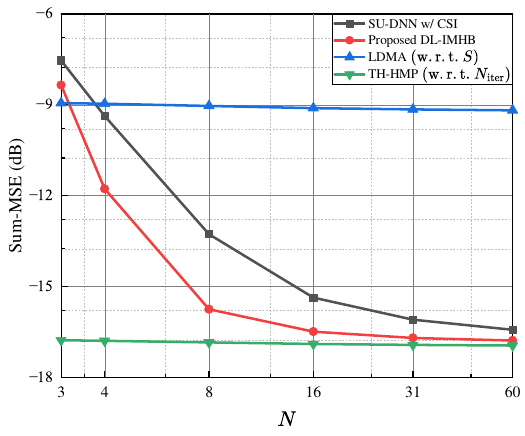}}
\caption{Performance evaluation in the PCSI-based indirect mode ($K=4$). The algorithm-specific complexity parameters are the number of virtual measurements $N$ for DL-IMHB and SU-DNN, distance rings $S$ for LDMA, and iterations $N_{\rm iter}$ for TH-HMP.}
\label{fig:pcsi_core_1}
\end{figure}

Fig.~\ref{fig:pcsi_core_1} reports the sum rate in Fig.~5(a) and the sum-MSE in Fig.~5(b) versus the algorithm-specific complexity parameters, namely the number of virtual measurements $N$ for DL-IMHB and SU-DNN, the number of distance rings $S$ for LDMA, and the iteration number $N_{\rm iter}$ for TH-HMP.
Among the learning-based methods, DL-IMHB consistently outperforms SU-DNN in both metrics. For example, at $N=8$, DL-IMHB achieves 29.42~bps/Hz, improving over SU-DNN by about 3.07~bps/Hz, and it also attains a lower sum-MSE of $-15.75$~dB versus $-13.28$~dB. This indicates that the variant-MMSE objective helps the analog-mapping network account for multiuser interference, especially when $N$ is small.
For the conventional baselines, LDMA saturates early with a much higher sum-MSE floor, which is consistent with its on-grid quantization. TH-HMP can approach similar performance with sufficiently many iterations, whereas it requires about $N_{\rm iter} \approx 10$ iterations to match the rate achieved by DL-IMHB at $N=8$, leading to higher latency under comparable settings as shown in Table~\ref{tab:complexity}.

\begin{figure}[!t]
\centering
\includegraphics[width=0.4\textwidth]{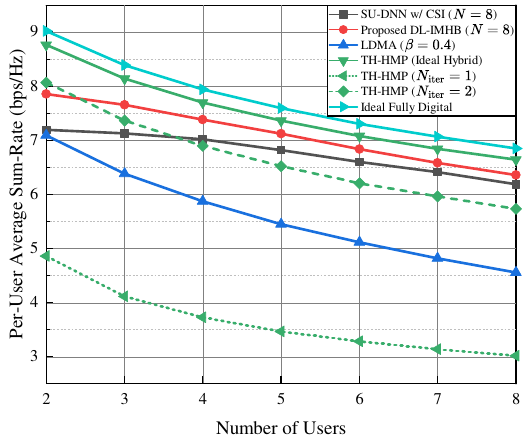}%
\caption{Scalability analysis regarding the number of users $K$ under the PCSI assumption ($N=8$). The baselines include LDMA ($\beta=0.4$) and TH-HMP under two distinct regimes: $N_{\rm iter}=200$ serves as the approximate performance upper bound, while $N_{\rm iter}\in\{1, 2\}$ represents complexity-constrained benchmarks comparable to the proposed scheme.}
\label{fig:pcsi_core_2}
\end{figure}

Fig. \ref{fig:pcsi_core_2} investigates the system scalability by presenting the achievable sum rate as a function of the user count $K$. 
To ensure a fair comparison in terms of computational load, the iterative TH-HMP algorithm is evaluated under a constrained low-latency setting $N_{\rm iter}\in\{1, 2\}$, alongside its fully converged but computationally prohibitive counterpart $N_{\rm iter}=200$. 
DL-IMHB maintains a clear sum rate advantage over the single-iteration TH-HMP and the codebook-based LDMA with $\beta=0.4$ over the considered user range. 
While the two-iteration TH-HMP exhibits marginal superiority at lower user loads, the proposed framework demonstrates superior scalability as the system density increases.
For $K=8$, DL-IMHB achieves 6.36 bps/Hz, compared with 5.73 bps/Hz of TH-HMP with $N_{\rm iter}=2$, while reducing the computational overhead by about $1.3\times$. 
When $K$ is large, a small iteration budget in TH-HMP is insufficient to suppress multiuser interference. In contrast, the proposed variant-MMSE formulation yields a KKT-derived closed-form digital precoder, and the merged output network enables coordinated beamforming across users.

Overall, the proposed framework scales well with the number of users and approaches near-optimal performance with a single forward pass, which is suitable for real-time indirect hybrid beamforming.

\subsection{Direct Hybrid Beamforming Performance}

\begin{figure}[!t]
\centering
\subfloat[Sum rate versus total pilot overhead at SNR=10~dB.]{%
  \includegraphics[width=0.4\textwidth]{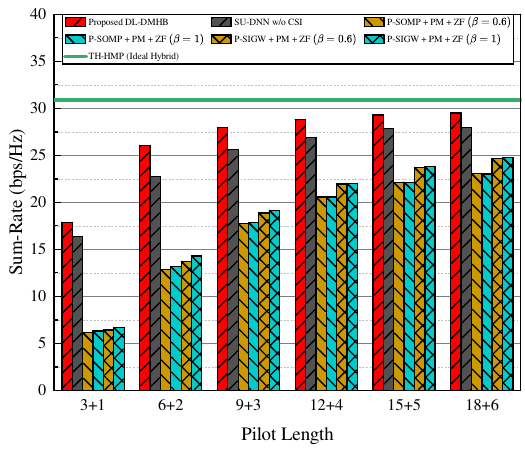}%
} \\
\subfloat[Sum rate versus SNR with different pilot configurations.]{%
  \includegraphics[width=0.4\textwidth]{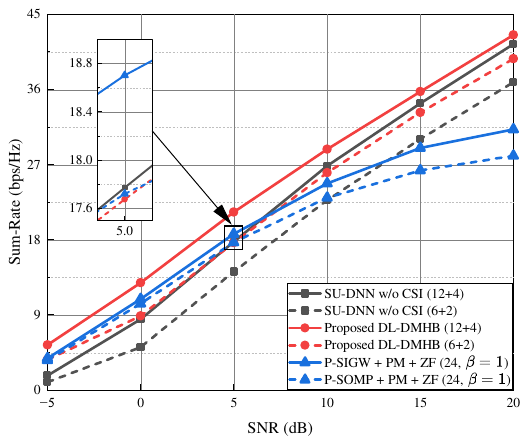}%
}
\caption{Performance evaluation of the direct beamforming schemes ($K=4, L=4$) without explicit CSI.}
\label{fig:ncsi_core_1}
\end{figure}

Fig.~\ref{fig:ncsi_core_1} compares direct beamforming schemes with respect to pilot efficiency in Fig.~\ref{fig:ncsi_core_1}(a) and SNR robustness in Fig.~\ref{fig:ncsi_core_1}(b).
In Fig.~\ref{fig:ncsi_core_1}(a), DL-DMHB attains the highest sum rate among the measurement-driven methods and gradually narrows the gap to the PCSI-aided TH-HMP benchmark as the pilot overhead $N{+}I$ increases. At the pilot-limited setting $N{+}I=6{+}2$, DL-DMHB achieves 26.07~bps/Hz, exceeding SU-DNN by 3.3~bps/Hz. In contrast, the sparse-recovery baselines exhibit notable degradation under the same pilot budget; for example, P-SIGW attains 14.29~bps/Hz, which is consistent with its sensitivity to grid mismatch.
Fig.~\ref{fig:ncsi_core_1}(b) shows that DL-DMHB maintains a consistent advantage over SU-DNN across the considered SNR range under fixed pilot budgets. Moreover, with $6{+}2$ pilots, DL-DMHB outperforms the 24-pilot P-SIGW baseline in the moderate-to-high SNR regime. For instance, at 10~dB, DL-DMHB achieves 26.07~bps/Hz, compared with 24.78~bps/Hz for P-SIGW. These observations suggest that the end-to-end design improves robustness under short pilots and mitigates the performance degradation associated with decoupled processing and sparse recovery.

\begin{figure}[!t]
\centering
\includegraphics[width=0.4\textwidth]{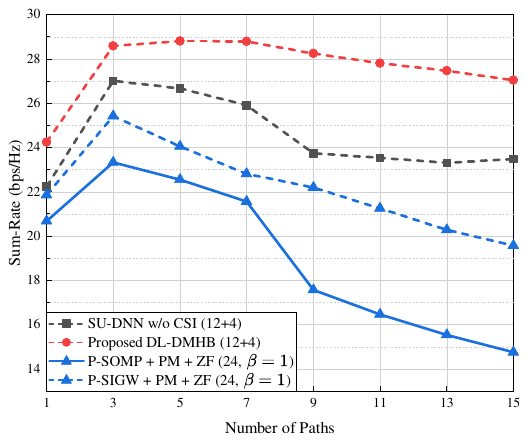}
\caption{Impact of channel sparsity on sum rate performance (SNR=10 dB, $K=4$), with the number of dominant paths $L$ varying from the sparse regime ($L=1$) to the scattering-rich regime ($L=15$). The deep learning methods are configured with $N=12$ sensing pilots and $I=4$ equivalent channel estimation pilots. }
\label{fig:perf_pcsi_numpaths}
\end{figure}

Fig.~\ref{fig:perf_pcsi_numpaths} examines the robustness of the proposed and baseline schemes with respect to channel sparsity by plotting the achievable sum rate versus the number of dominant paths $L$.
In the sparse regime with $L \le 3$, all evaluated methods benefit from increased multipath diversity and exhibit rising sum rates. As the channel becomes scattering-rich with larger $L$, the algorithms diverge in performance due to their different modeling assumptions.
P-SOMP degrades rapidly beyond $L=7$, which is consistent with grid mismatch and the breakdown of sparsity assumptions. P-SIGW exhibits a more gradual decline, since off-grid refinement partially alleviates the mismatch.
The SU-DNN baseline also encounters a performance bottleneck around $L=9$ and then stabilizes in the high-$L$ regime, as it relies on statistical feature extraction rather than explicit path decomposition.
With $N=12$ and $I=4$, DL-DMHB maintains 27.03~bps/Hz at $L=15$, which is about 3.5~bps/Hz higher than SU-DNN.

\begin{figure}[!t]
\centering
\includegraphics[width=0.4\textwidth]{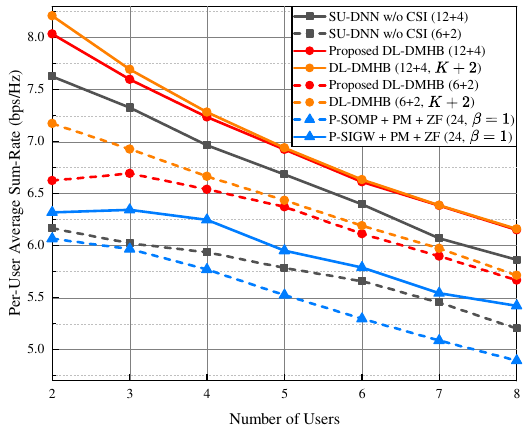}
\caption{Scalability analysis of direct beamforming schemes regarding the number of users $K$ (SNR=10 dB, $L=4$). The evaluation includes the standard configuration $N_{\rm RF}=K$ and the relaxed hardware constraint $N_{\rm RF}=K+2$ to assess the exploitation of excess spatial degrees of freedom.}
\label{fig:perf_pcsi_numusers}
\end{figure}

Fig.~\ref{fig:perf_pcsi_numusers} plots the per-user average sum rate as the number of users increases from 2 to 8. As the user load increases, the per-user rate decreases due to stronger multiuser interference, which is consistent with the indirect beamforming results.
The SU-DNN architecture requires the number of RF chains to match the number of users in order to maintain fixed network dimensions, and therefore cannot be evaluated when redundant RF chains are available. In contrast, DL-DMHB can be applied with more RF chains than users since it operates directly on the effective measurements.
With a pilot budget of $6{+}2$ and 2 active users, increasing the number of RF chains to $N_{\rm RF}=K{+}2$ raises the per-user rate from 6.63 to 7.17~bps/Hz.
Under the standard configuration with $N_{\rm RF}=K$ at 8 users, the proposed $6{+}2$ scheme achieves a per-user rate of 5.67~bps/Hz, which remains higher than that of the 24-pilot P-SIGW baseline at 4.89~bps/Hz.

\begin{figure}[!t]
\centering
\includegraphics[width=0.4\textwidth]{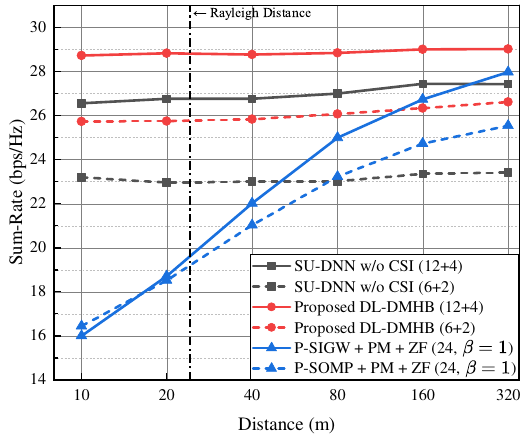}
\caption{Performance robustness analysis across near-to-far field propagation regimes (SNR=10 dB, $K=4, L=4$). The achievable sum rate is plotted against the maximum path distance, ranging from the extreme near-field ($10$~m) to the far-field limit ($320$~m).}
\label{fig:perf_pcsi_dist}
\end{figure}

Fig.~\ref{fig:perf_pcsi_dist} examines robustness with respect to user distribution ranges by varying the maximum path distance $r_{\max}$ from 10~m to 320~m, covering extreme near-field to mixed near- and far-field conditions.
The results indicate that explicit channel reconstruction methods are increasingly challenged in very short-range scenarios, whereas DL-DMHB maintains stable performance over the entire distance range.
At $r_{\max}=10$~m, the proposed $12{+}4$ configuration achieves 28.72~bps/Hz, exceeding the SU-DNN baseline by more than 2~bps/Hz.
In contrast, the polar-domain sparse recovery baselines, including P-SOMP and P-SIGW, exhibit pronounced degradation in the extreme near-field. For instance, at $r_{\max}=10$~m, the 24-pilot P-SIGW attains 16.01~bps/Hz, whereas the proposed $6{+}2$ configuration achieves 25.73~bps/Hz. This trend is consistent with the breakdown of sparsity assumptions caused by severe energy spread at short distances.
An anomaly is observed at $r_{\max}=10$~m, where P-SIGW performs slightly worse than P-SOMP, suggesting instability of gradient-based distance refinement under strong near-field curvature.
As $r_{\max}$ increases toward the planar-wave regime, the sparse recovery baselines gradually recover performance; however, DL-DMHB consistently preserves a clear advantage while requiring substantially fewer pilot resources.

\begin{figure}[!t]
\centering
\includegraphics[width=0.4\textwidth]{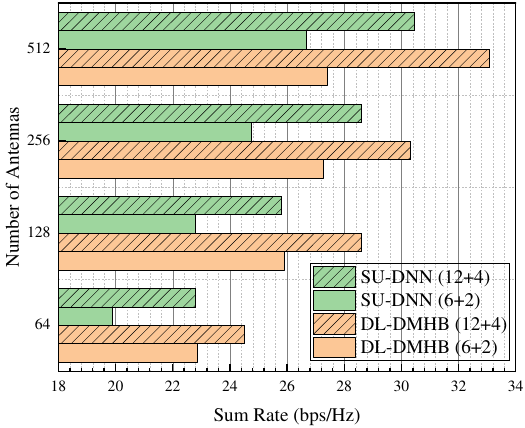}
\caption{Achievable sum rate performance under different uniform planar array (UPA) configurations ($M_y \times M_z$) with varying pilot overheads (SNR=10 dB, $K=4$). The total number of antennas $M$ ranges from 64 ($8\times8$) to 512 ($32\times16$).}
\label{fig:upa_scaling}
\end{figure}

Finally, we investigate the scalability of the proposed framework and its generalization to practical XL-MIMO array geometries by extending the evaluation to uniform planar arrays (UPAs). 
Fig.~\ref{fig:upa_scaling} illustrates the sum-rate performance across various antenna configurations, ranging from a small-scale $8\times8$ array to a large-scale $32\times16$ array.
The proposed DL-DMHB adapts effectively to the UPA structure without requiring architectural modifications, demonstrating robustness against changes in array geometry beyond the standard ULA.
As the number of antennas $M$ scales up to 512, the system performance improves monotonically due to the increased array gain. 
Notably, even at $M=512$, the proposed model converges stably and achieves a sum rate of 27.40 bps/Hz with reduced overhead with $6+2$ pilots, outperforming the 26.67 bps/Hz of the SU-DNN baseline. 
While the performance gap narrows slightly at minimal pilot overhead due to array gain-induced capacity saturation, DL-DMHB expands its lead to 2.6 bps/Hz under the $12+4$ setting.

\subsection{Interpretability of Learned Sensing and Features}
\label{subsec:interpretability}
To provide insights into what the proposed network learns beyond E2E performance, we visualize the effective sensing patterns in the initial UL sensing phase and the learned complex-valued features extracted by the shared per-user multiplexing module.

\begin{figure*}[t]
    \centering
    \includegraphics[width=0.4225\textwidth]{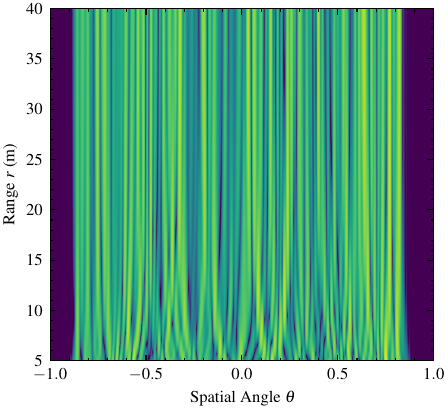} 
    \includegraphics[width=0.435\textwidth]{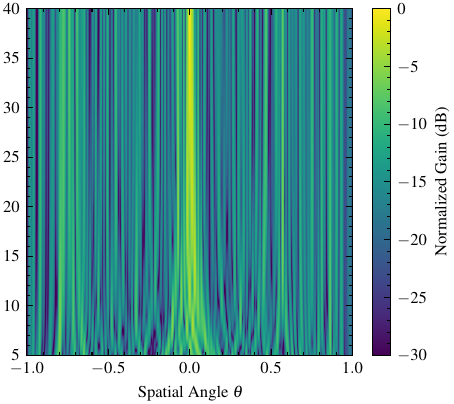}
    \caption{Representative sensing pattern visualization via codebook correlation. Left: DL-DMHB with a grouped-conv kernel interpreted as an effective sensing vector. Right: SU-DNN with an RF-chain sensing vector from the phase-only linear layer.}
    \label{fig:perception_beams}
\end{figure*}

We first interpret the learned weights as effective sensing vectors. 
It is important to note the structural distinction between the two compared schemes. 
For the proposed DL-DMHB, the effective sensing matrix is synthesized through the interplay of analog RF sensing and digital baseband aggregation, creating unconstrained complex-valued weights. 
In contrast, the SU-DNN baseline relies on a phase-only analog linear layer. 
By correlating these vectors with a near-field spherical-wave codebook over the angular domain $\theta\in[-1,1]$ and range domain $r\in[5,40]$~m, we obtain the spatial response profiles shown in Fig.~\ref{fig:perception_beams}. 
The SU-DNN baseline exhibits a rigid sensing behavior where the energy is predominantly concentrated around the broadside direction at $\theta \approx 0$. 
Further analysis reveals that this baseline scheme lacks beam diversity and tends to focus redundantly on a fixed direction rather than performing spatial scanning. 
Conversely, the proposed DL-DMHB generates a multi-lobed pattern with broader angular coverage. 
A notable observation is that the sensing energy is suppressed at the extreme angular edges $|\theta| \approx 1$. 
This indicates that the network learns to concentrate sensing resources within the effective sector where the array gain is significant, while avoiding energy wastage in grazing angle directions where the effective aperture is reduced.

\begin{figure}[!t]
    \centering
    \includegraphics[width=0.45\textwidth]{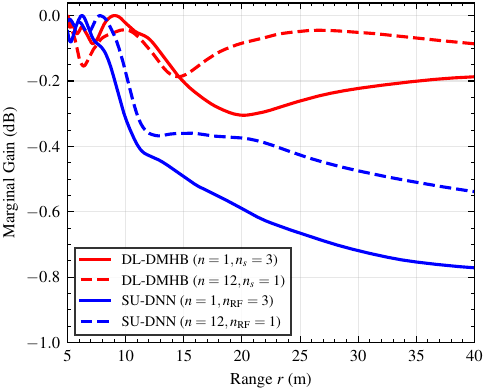}
    \caption{Comparison of the normalized range-marginal gain profiles. The curves illustrate the sensing sensitivity distribution along the range dimension for selected measurement slots ($n$), and data streams ($n_s$) or RF chains ($n_{\rm RF}$), computed by integrating the beam response across the angular domain.}
    \label{fig:gains}
\end{figure}

We further evaluate range-dependent sensing by integrating the beam response over $\theta$ to obtain the range-marginal profile in Fig.~\ref{fig:gains}. SU-DNN shows a clear decay in marginal gain as $r$ increases, while DL-DMHB maintains a flatter profile over the same range. This indicates that the learned sensing front-end generalizes better across near- and far-field wavefronts. The subsequent user separation is handled by the features extracted by the multiplexing module.

\begin{figure}[!t]
\centering
\subfloat[Distance-Conditioned PC1 Phase Manifold]{%
  \includegraphics[width=0.4\textwidth]{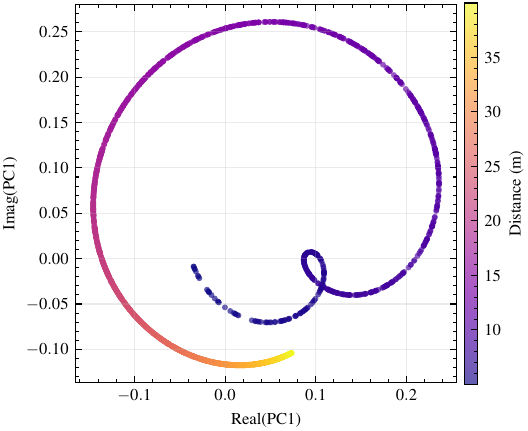}%
} \\
\subfloat[Distance-Conditioned Magnitude Embedding]{%
  \includegraphics[width=0.4\textwidth]{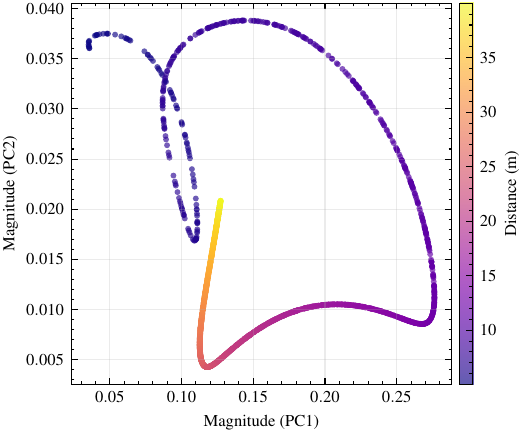}%
}
\caption{Visualization of the shared-MLP feature manifold using complex PCA.}
\label{fig:mlp_cpca}
\end{figure}

Finally, we probe the representations learned by the shared MLP in the multiplexing module using complex principal component analysis (PCA). We conduct a controlled experiment using a single-path spherical-wave channel with a fixed angle $\theta=0$ and uniformly sampled distance. Crucially, the channel complex gain is normalized to isolate the structural changes in the steering vector $\mathbf{b}(\theta, r)$. Fig.~\ref{fig:mlp_cpca} visualizes the first complex principal component (PC1) in the complex plane and the embedding in the magnitude space. As shown in Fig.~\ref{fig:mlp_cpca}(a), the projected feature points form a smooth arc-like trajectory, which varies smoothly with the near-field phase evolution as $r$ changes. As the distance $r$ increases, the quadratic phase term across the antenna array proportional to $1/r$ diminishes, causing the channel vector to rotate continuously in the high-dimensional observation space. The network captures this phase evolution and maps it to a continuous manifold in the latent space. Furthermore, Fig.~\ref{fig:mlp_cpca}(b) illustrates the non-linear relationship between the magnitudes of the first two principal components. The varying projection magnitude along PC1 indicates that the geometry of the channel manifold is curved. These results suggest that the learned features capture range-dependent phase characteristics, which helps distinguish users with different wavefront curvatures.

\subsection{Discussion and Future Work}
\subsubsection{Extension to Wideband OFDM and Beam Squint}
Although this work assumes a narrowband channel, the proposed framework is theoretically scalable to wideband OFDM systems.
In hybrid architectures, the analog precoder $\mathbf F_{\rm RF}$ is constrained to be frequency-flat.
Consequently, the proposed neural network architecture requires no structural modification to the output layer, as it predicts a single $\mathbf F_{\rm RF}$ shared across all $N_{sc}$ subcarriers.
The frequency selectivity and beam squint effects are managed by the digital precoder, which is computed per-subcarrier using the derived KKT closed-form solution as $\tilde{\mathbf F}_{\rm BB}[n] = (\mathbf H_{\rm eq}^H[n]\mathbf H_{\rm eq}[n]+\mu\mathbf F_{\rm RF}^H\mathbf F_{\rm RF})^{-1}\mathbf H_{\rm eq}^H[n]$.
During training, the loss function can be formulated as the sum-rate averaged over subcarriers.
By optimizing this aggregated objective, the network learns a robust $\mathbf F_{\rm RF}$ that balances the beam squint effect across the bandwidth, avoiding the high complexity of per-subcarrier neural network processing.

\subsubsection{Practical Hardware Impairments}
While this work assumes ideal hardware, practical deployment faces challenges such as quantized phase shifters, power amplifier (PA) nonlinearities, and RF calibration errors. The proposed framework can be extended to support quantized phase constraints by incorporating differentiable quantization layers such as Gumbel-Softmax during training. Regarding PA efficiency, our strict enforcement of the CM constraint helps mitigate non-linear distortions by reducing the peak-to-average power ratio (PAPR) of the analog signals. Furthermore, for RF calibration errors, DL-DMHB offers a promising solution, that is, by training on datasets capable of capturing hardware mismatches via over-the-air measurements, the E2E network can implicitly learn to compensate for channel-hardware impairments jointly. These practical extensions will be the subject of our future work.

\section{Conclusion}
This paper presents a fully complex-valued E2E framework for multiuser hybrid beamforming in near- and far-field XL-MIMO systems.
The network includes a grouped complex-convolution sensing front-end, a shared complex MLP, and a merged output head with CM normalization.
Training minimizes a variant-MMSE objective where the digital precoder is eliminated in closed form via KKT conditions.
This decoupling improves optimization stability and keeps the learned analog precoder consistent with the hardware constraints.
The proposed E2E network seamlessly unifies both ``indirect'' and ``direct'' operational modes. Particularly in the direct mode, the learned sensing operator and analog mapper facilitate robust short-pilot acquisition, which is subsequently complemented by efficient equivalent-channel estimation and a closed-form digital precoder.
Numerical simulations show that the proposed approach consistently outperforms conventional sparse-recovery pipelines and representative baselines across diverse near- and far-field propagation scenarios.
Specifically, it delivers substantial sum-rate gains of up to 3 bps/Hz and demonstrates superior pilot efficiency, all while closely approaching the theoretical performance upper bounds of fully digital and ideal hybrid beamforming with low inference complexity.


\bibliographystyle{IEEEtran}
\bibliography{IEEEabrv,References}

\vfill

\end{document}